%% file: main.tex
\documentclass[manuscript, screen]{acmart}

\AtBeginDocument{%
  }

\setcopyright{acmlicensed}
\copyrightyear{2025}
\acmYear{2025}
\acmDOI{XXXXXXX.XXXXXXX}

\acmISBN{978-1-4503-XXXX-X/18/06}

\usepackage{algorithm}
\usepackage{algpseudocode}
\usepackage{url}
\usepackage{amsmath}
\usepackage{graphicx}
\usepackage{tabularx}
\usepackage{multirow}
\usepackage{threeparttable}
\usepackage{colortbl}
\usepackage{tcolorbox}
\usepackage{adjustbox}
\usepackage{booktabs}  
\usepackage{arydshln}  

\definecolor{codegreen}{rgb}{0,0.6,0}
\definecolor{codegray}{rgb}{0.5,0.5,0.5}
\definecolor{codepurple}{rgb}{0.58,0,0.82}
\definecolor{backcolour}{rgb}{0.95,0.95,0.92}
\definecolor{lightgray}{gray}{0.95}

\begin{document}

\title{SETBVE: Quality-Diversity Driven Exploration of Software Boundary Behaviors}


\author{Sabinakhon Akbarova}
\orcid{0000-0002-6518-8193}
\email{sabina.akbarova@chalmers.se}
\affiliation{%
  \institution{Department of Computer Science and Engineering, 
  Chalmers University of Technology and University of Gothenburg}
  \city{Gothenburg}
  \country{Sweden}
}

\author{Felix Dobslaw}
\orcid{0000-0001-9372-3416}
\email{felix.dobslaw@miun.se}
\affiliation{%
  \institution{Department of Communication, Quality Management and Information Systems,  
  Mid Sweden University}
  \city{Östersund}
  \country{Sweden}
}

\author{Francisco Gomes de Oliveira Neto}
\orcid{0000-0001-9226-5417}
\email{francisco.gomes@cse.gu.se}
\affiliation{%
  \institution{Department of Computer Science and Engineering, 
  Chalmers University of Technology and University of Gothenburg}
  \city{Gothenburg}
  \country{Sweden}
}

\author{Robert Feldt}
\orcid{0000-0002-5179-4205}
\authornote{Corresponding author}
\email{robert.feldt@chalmers.se}
\affiliation{%
  \institution{Department of Computer Science and Engineering, 
  Chalmers University of Technology and University of Gothenburg}
  \city{Gothenburg}
  \country{Sweden}
}

\renewcommand{\shortauthors}{Akbarova et al.}

\begin{abstract}
Software systems exhibit distinct behaviors based on input characteristics, and failures often occur at the boundaries between input domains. Traditional Boundary Value Analysis (BVA) relies on manual heuristics, while automated Boundary Value Exploration (BVE) methods typically optimize a single quality metric, risking a narrow and incomplete survey of boundary behaviors. We introduce SETBVE, a customizable, modular framework for automated black-box BVE that leverages Quality-Diversity (QD) optimization to systematically uncover and refine a broader spectrum of boundaries. SETBVE maintains an archive of boundary pairs organized by input- and output-based behavioral descriptors. It steers exploration toward underrepresented regions while preserving high-quality boundary pairs and applies local search to refine candidate boundaries. In experiments with ten integer‐based functions, SETBVE outperforms the baseline in diversity, boosting archive coverage by 37 to 82 percentage points. A qualitative analysis reveals that SETBVE identifies boundary candidates the baseline misses. While the baseline method typically plateaus in both diversity and quality after 30 seconds, SETBVE continues to improve in 600-second runs, demonstrating better scalability. Even the simplest SETBVE configurations perform well in identifying diverse boundary behaviors. Our findings indicate that balancing quality with behavioral diversity can help identify more software edge-case behaviors than quality-focused approaches.
\end{abstract}




\maketitle

\input{files/introduction}
\input{files/background_rw}
\input{files/approach}
\input{files/methodology}

\input{files/results}
\input{files/discussion}

\input{files/conclusion}

\begin{acks} 
This work was supported by the Wallenberg AI, Autonomous Systems and Software Program (WASP) funded by the Knut and Alice Wallenberg Foundation. Robert Feldt and Felix Dobslaw have also been supported by the Swedish Scientific Council (No. 2020-05272, `Automated boundary testing for Quality of AI/ML models'). ChatGPT and Claude.ai was utilized to improve the phrasing of some parts of the text, originally written by the authors.
\end{acks}

%
\bibliographystyle{ACM-Reference-Format}
\bibliography{references}


\end{document}

%% file: files/introduction.tex
\section{Introduction}

Software testing plays a crucial role in ensuring software quality, primarily through the selection of test data that evaluates whether a program behaves as intended. Ideally, testing would cover all possible inputs to verify that software functions correctly. However, covering all inputs is generally infeasible due to the vast input space in realistic software systems. For example, consider a \texttt{Date} class, where testing every combination of day, month, and year would result in an enormous number of cases. Consequently, the main challenge in software testing is selecting test cases that are diverse enough to capture a broad and representative range of program behaviors.

To address this challenge, a variety of test selection methods have been developed, aiming to identify representative test cases. One category of these methods is input domain-based techniques, which focus on how test cases are generated from the input space. Among these techniques, Boundary Value Analysis (BVA) and testing analyze software artifacts to identify the discrepancies between expected and actual boundaries ~\cite{myers2011art, cohen1980domain, clarke1982aclose}. Hierons \cite{hierons2006avoiding} defined boundaries as \textit{pairs of inputs} that are close to each other but fall within adjacent sub-domains, where these sub-domains represent distinct behavior domains in the context of a software under test (SUT). BVA is widely used for its effectiveness in detecting faults as they tend to occur near boundaries \cite{bourque2014swebok}. However, identifying boundaries within the input space remains challenging due to the lack of clear, objective methods, and even Grochtmann and Grimm ~\cite{grochtmann1993classification} noted that BVA is a creative, manual process that cannot be automated.

\begin{figure}[h]
    \centering
    \includegraphics[width=\textwidth]{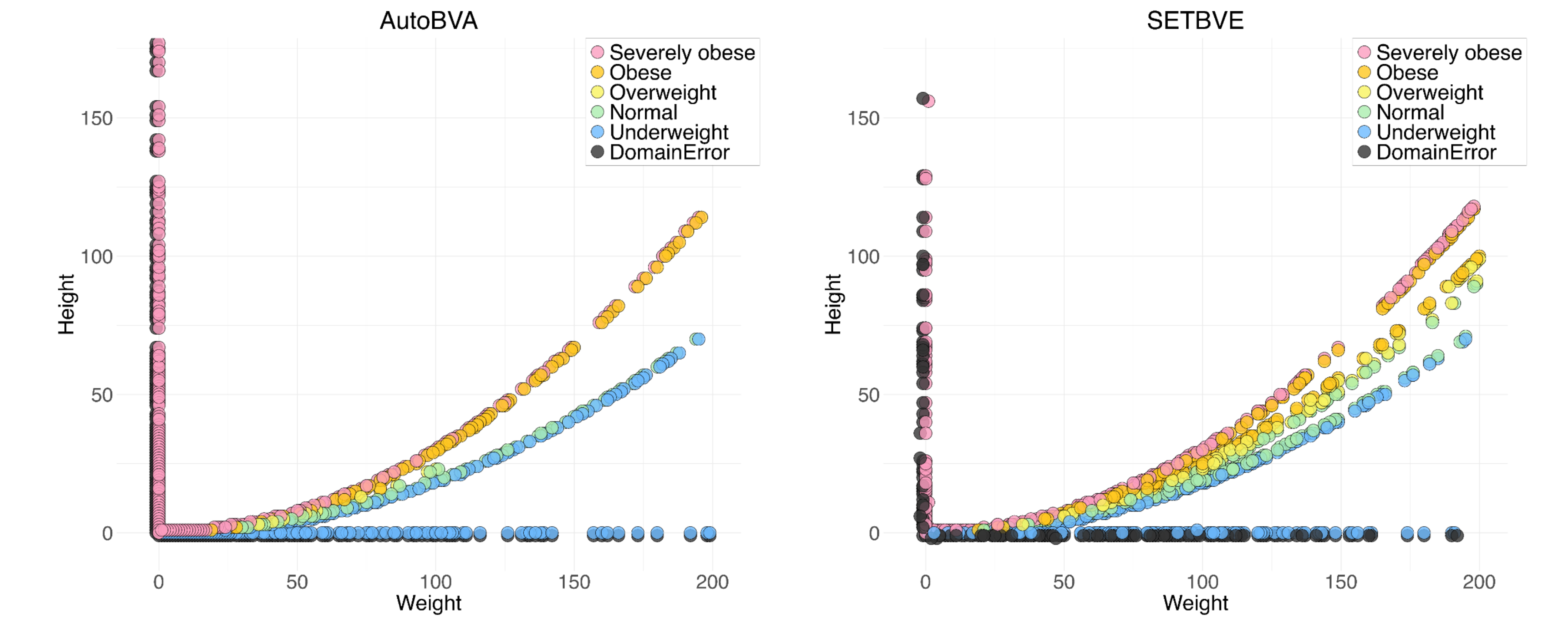}
    \caption{Motivating example comparing BMI boundaries found by AutoBVA and SETBVE.}
    \label{fig:motivating_example} 
\end{figure}

Challenging this assumption, Dobslaw et al.~\cite{dobslaw2020bve} introduced the concept of Boundary Value Exploration (BVE), a set of methods for systematically identifying input pairs that lie on opposite sides of a behavioral boundary. They proposed AutoBVA, the first automated black-box framework for boundary detection, shifting boundary analysis away from its traditionally manual nature\cite{dobslaw2023automated}. AutoBVA effectively discovers \textit{boundary candidates}—input pairs that are close in the input space but yield distinct outputs—and ranks these candidates based on their program derivative, a metric designed to capture the sensitivity of software behavior to input changes~\cite{feldt2019PD}. However, the BVE task is complicated by the presence of multiple boundaries, each potentially having a distinct maximum program derivative~\cite{feldt2019PD}. Consequently, focusing solely on maximizing program derivative, as AutoBVA does, may limit exploration to a narrower set of boundary candidates, concentrating the discovered boundaries into fewer behavioral regions. This approach potentially overlooks broader behavioral diversity and misses boundary candidates that, while exhibiting lower derivative values, are nonetheless important for comprehensive software testing.

This is where Quality-Diversity (QD) optimization offers a promising alternative. Unlike traditional optimization methods that seek a single best solution, QD aims to discover a broad set of diverse, high-performing solutions~\cite{chatzilygeroudis2021quality}. Its success across domains like robotics~\cite{mouret2015illuminating, cully2015robots, cully2016evolving, duarte2017evolution}  and video games ~\cite{liapis2015constrained, gravina2019procedural, canaan2019diverse, ecoffet2021first} has recently extended into software testing, including applications in both traditional ~\cite{boussaa2015noveltysearch, marculescu2016using, feldt2017searching} and deep learning-based systems ~\cite{riccio2020model, zohdinasab2021deephyperion, zohdinasab2023efficient}. Given its dual focus on solution quality and diversity, QD provides a natural fit for the challenges of BVE. We argue that QD is particularly well suited for BVE for at least two main reasons. First, it supports a customizable \textit{behavioral space} --- also referred to as feature space --- allowing testers to tailor exploration objectives to specific testing goals. For instance, one tester may aim to broadly cover different input characteristics, while another may focus on fewer input features but seek maximum diversity in output behavior. This flexibility enables adaptive exploration and supports varied testing strategies. Second, QD algorithms inherently promote diversity by encouraging the exploration of multiple regions in the behavioral space. The emphasis on diversity makes QD especially valuable in boundary exploration, as it helps uncover a wider variety of boundaries and supports more comprehensive coverage of the behavioral space.

Despite the growing interest in QD-based methods in software testing, to the best of our knowledge, BVE has not yet been formulated as a QD optimization problem. In this paper, we introduce SETBVE, a novel framework for automatically discovering diverse boundary behaviors in SUTs. SETBVE incorporates QD algorithms to explore the input space more comprehensively and identify uncover a wider range of boundary candidates. In contrast, AutoBVA focuses on maximizing boundary candidate quality through the program derivative, which can lead to a narrow search that overlooks regions with lower derivative values and often yields only a few high-quality boundary candidates. 
A motivating example can be seen in Figure ~\ref{fig:motivating_example}, consider a Body Mass Index (BMI) function that classifies input pairs (height, weight) into distinct categories such as ``Underweight'', ``Normal'', and ``Obese''. The results in the Figure reveals that SETBVE discovers a broader set of BMI boundaries, capturing transitions between diverse behaviors that AutoBVA misses. This simple example highlights the potential of QD optimization to uncover more diverse boundary regions, supporting more thorough boundary exploration.

SETBVE is built around three components: \textbf{S}ampler, \textbf{E}xplorer, and \textbf{T}racer. The modular design of SETBVE allows users to combine and configure those components flexibly according to testing goals. Each component plays a distinct role and can be adapted independently. The Sampler sets up an archive --- a grid-like structure that stores candidates based on selected properties. These properties, called \textit{behavioural descriptors}, capture key aspects of the SUT, such as input features or output responses. In this work, we define descriptors using both input and output characteristics, but they can be customized to suit different testing needs. Explorer navigates the input space by selecting existing archive entries and mutating them to discover new archive cells. Tracer further refines the search by examining areas around the boundary candidates found by the Sampler and/or Explorer. It aims to follow detected boundary lines and uncover additional boundary candidates nearby. 

Similar to AutoBVA, SETBVE runs fully automatically, requiring no access to source code or formal specifications. The current implementation supports any number of integer inputs and handles outputs by stringifying them, allowing for compatibility with both numeric and textual output formats. With modifications, such as adjusting the mutation operator or generalizing quality metrics~\cite{feldt2019PD}, the framework can be extended to support other input and output types as well. To support further research and ensure reproducibility, we provide the full implementation of SETBVE along with all scripts needed to replicate the experimental evaluation\footnote{\url{https://github.com/aksabina/SETBVE}}, as well as the associated dataset\footnote{\url{https://zenodo.org/records/15364606}} .

We evaluate SETBVE by examining the performance of its various configurations, including a comparison with AutoBVA. The comparison is conducted through experiments on ten different systems under test (SUTs). For each technique, we assess the quality and diversity of the identified boundary candidates, as well as the trade-off between these two aspects. Our results suggest that QD algorithms offer clear advantages for BVE, particularly by increasing the diversity of identified boundary candidates. SETBVE consistently identifies more diverse candidates than AutoBVA while still maintaining relatively high-quality results, especially as the SUT complexity increases. Depending on the characteristics of a SUT, the Tracer component further contributes by identifying additional boundary candidates near those already discovered, leading to a more complete and interpretable view of behavioral transitions.

In this paper, we make the following contributions:
\begin{itemize}
\item We propose SETBVE, a customizable and modular framework for automated boundary value exploration that identifies boundary candidate pairs by incorporating ideas from Quality-Diversity optimization.
\item We empirically evaluate multiple configurations of SETBVE across ten SUTs, demonstrating its effectiveness in identifying diverse and high-quality boundary candidates, outperforming existing methods ~\cite{dobslaw2023automated}.
\item For the empirical comparison, we define BVE-specific metrics --- based on related QD literature~\cite{mouret2015illuminating} --- to assess the quality and behavioral spread of boundary candidates.
\item We introduce a Tracer component that explores the vicinity of identified boundary candidates, delineating additional boundary transitions and refining the representation of boundary regions.
\end{itemize}

The rest of the article is organized as follows. In Section \ref{sec:background}, we provide background information and discuss related work. Section \ref{sec:approach} then presents our proposed method, SETBVE, in detail. Section \ref{sec:methodology} outlines the methodology of our evaluation. Section \ref{sec:results} covers the results of our empirical evaluation and responses to our research questions. In Section \ref{sec:discussion}, we discuss the study’s implications and address potential threats to validity. Finally, Section \ref{sec:conclusion} concludes the paper.

%% file: files/background_rw.tex
\section{Background and related work}
\label{sec:background}

In this section, we provide a brief background and review relevant studies on Boundary Value Analysis, beginning with traditional boundary testing concepts, followed by recent advancements in automated frameworks. Finally, we present an overview of Quality-Diversity optimization methods and their application in software testing.

\subsection{Boundary Value Analysis}

Boundary Value Analysis (BVA) is a fundamental and widely adopted technique in software testing, focusing on inputs at the edges of input domains where errors are most likely to occur~\cite{clarke1982aclose,hierons2006avoiding}. White and Cohen~\cite{cohen1980domain} proposed a domain-testing strategy focused on identifying boundaries between mutually exclusive subdomains of the input space to detect control-flow errors. To improve the efficiency of testing boundaries, Jeng et al.~\cite{jeng1999automatic} introduced a semi-automated method that combines dynamic search with algebraic manipulation of boundary conditions. 

Even with its well-established role in software testing, BVA continues to be an active area of research, with recent studies exploring its integration with modern techniques such as search-based software testing. Ali et al.~\cite{ali2016generating} extended a search-based test data generation technique to model-based testing by integrating a solver that generates boundary values using heuristic guidance. However, to advance BVA in modern software systems, existing approaches need to deepen their understanding of what constitutes a boundary and how such boundaries can be effectively identified from a black-box perspective.

More recently, Dobslaw et al.~\cite{dobslaw2020bve} introduced Boundary Value Exploration (BVE), a concept designed to complement traditional BVA by supporting boundary detection in cases where specifications are incomplete or missing. BVE employs distance-based metrics to systematically explore the input space and quantify behavioral changes, enabling the identification of boundary regions more effectively.

One way of quantifying such boundaries is the concept of a \textit{program derivative} (PD), introduced by Feldt et al. ~\cite{feldt2019PD}. Program derivative, inspired by the mathematical derivative of a function, serves as a measure of boundariness by quantifying the sensitivity of a program’s behavior (output) to changes in its input. Formally, given two input values $i_1$ and $i_2$, and their corresponding outputs $o_1$ and $o_2$, the PD is calculated as the ratio of the output distance to the input distance:

\begin{equation}
    PD(i_1, i_2) = \frac{d_o(o_1, o_2)}{d_i(i_1, i_2)}
\end{equation}

Here, $d_i(i_1, i_2)$ is a distance measure on the input space, representing the difference between the inputs $i_1$ and $i_2$, whereas $d_o(o_1, o_2)$ is a distance measure on the output space, capturing how much the program output changes in response. An input pair is considered a \textit{boundary candidate} if it has a PD greater than zero, meaning the inputs are close but produce different outputs. A high PD indicates a strong boundary region (i.e. high boundariness), where small variations in input result in significant changes in the program’s behavior, making PD particularly valuable for BVE.

Traditionally, boundaries are tightly coupled with the specific behavior of a system under test (SUT), making them highly system-dependent; this presents a challenge in developing more general boundary definitions that can be applied across diverse SUTs. Therefore, based on the concept of boundary candidates, Dobslaw et al.~\cite{dobslaw2023automated} introduced the notion of \textit{validity groups}, which categorizes each boundary candidate --- consisting of two inputs and their corresponding outputs --- into one of three types: VV, where both outputs are valid; VE, where one output is valid and the other is an error; and EE, where both outputs are errors. These concepts have made it possible to define quality goals (e.g., measured by the PD) and, by systematically varying inputs, to explore the input space and automate boundary identification.

\subsubsection{Automation of Boundary Value Analysis}

Automated methods for testing partitions and boundary values commonly depend on available software specifications. For example, Pandita et al. ~\cite{pandita2010guided} proposed a white-box testing approach that enhances boundary coverage by instrumenting the SUT to identify boundary values near existing decision points. Zhang et al. \cite{zhang2015bvawhitebox} proposed a BVA-aware white-box method to improve fault coverage by identifying and testing boundaries derived from comparison predicates in symbolic execution paths. They introduced constrained combinatorial testing to generate test cases, covering boundary conditions with fewer tests while maintaining structural coverage. Hübner et al. ~\cite{hubner2019experimental} developed an equivalence class partitioning (ECP) strategy to efficiently locate boundary regions between equivalence classes. Guo et al. ~\cite{guo2023towardshigh} use machine learning with inputs and execution paths to learn input boundaries and apply Markov chain Monte Carlo (MCMC) to generate test cases. While effective in some scenarios, those white-box approaches heavily rely on clearly defined partitions or program specifications, limiting their applicability or effectiveness when specifications are ambiguous, incomplete, or unavailable.

To address some of those limitations, Dobslaw et al. introduced AutoBVA, an automated black-box boundary value exploration framework~\cite{dobslaw2023automated}. Their study demonstrated that black-box BVA could be automated, with AutoBVA successfully identifying notable boundary candidates. AutoBVA operates in two primary phases: detection and summarization. In the detection phase, the algorithm searches the input space to discover potential boundary candidates. The authors experimented with two alternative search strategies: Local Neighbourhood Search (LNS) and Boundary Crossing Search (BCS), with BCS yielding better results. Once the detection phase is complete, the summarization phase clusters the identified boundary candidates to provide a concise summary of boundary candidates for testers. Interestingly, during their evaluation, the authors uncovered unexpected behavior in one of Julia’s base functions, which led to raising a GitHub issue\footnote{\url{https://github.com/JuliaLang/julia/issues/48971}} and later contributing a documentation fix\footnote{\url{https://github.com/JuliaLang/julia/commit/e4c90e22e999e85268fc5465b2840df6f4f2fb94}} --- highlighting that BVE can reveal subtle edge cases even in well-established standard libraries.

Dobslaw et al. ~\cite{dobslaw2023automated} also introduced a sampling approach that combines \textit{compatible type sampling} (CTS) with bituniform sampling. Bituniform sampling randomly selects numbers using bit-shifting to insert leading zeros, ensuring broad exploratory coverage. CTS complements this by sampling argument-wise based on compatible data types --- for instance, integer types of different bit sizes, such as booleans (\texttt{Bool}), 8-bit integers (\texttt{Int8}), and 32-bit integers (\texttt{Int32})\footnote{In Julia programming language}, are compatible as they share the integer supertype. Their experiments showed that this combination outperformed other sampling strategies.

Despite AutoBVA’s strengths, the method revealed challenging research gaps. Due to the focus on high-quality boundary candidates, the framework may miss certain regions within the input space, resulting in reduced diversity of discovered solutions. Additionally, its computationally intensive summarization phase can introduce overhead, especially when analyzing complex SUTs. Our proposed framework, SETBVE, aims to overcome those barriers by leveraging on quality diversity optimization.

\subsection{Quality Diversity Optimization}
Quality-Diversity (QD) optimization algorithms stand apart from typical stochastic optimization approaches by seeking a broad range of high-performing solutions across a feature space rather than a single optimum~\cite{chatzilygeroudis2021quality}. This feature space, often called the \textit{behavioral space}, reflects the various behaviors of candidate solutions. Unlike multimodal optimization, which targets multiple optima, QD aims to cover the entire behavioral space, providing a wide variety of viable solutions. This breadth makes QD algorithms particularly valuable in domains where diverse solution behaviors are essential. However, these methods can face challenges in high-dimensional, noisy environments and can be computationally intensive, especially when fine-grained detail across the behavioral space is necessary~\cite{chatzilygeroudis2021quality}.

Early QD algorithms include novelty search~\cite{lehman2011abandoning} and MAP-Elites~\cite{mouret2015illuminating}. Novelty search, introduced by Lehman and Stanley, encourages exploration by rewarding new behaviors~\cite{lehman2011abandoning}. It assesses novelty by measuring the distance to the closest observed solutions, promoting clusters of unique solutions without necessarily spreading evenly across all behavioral features. In contrast, MAP-Elites by Mouret and Clune uses an illumination-based approach to systematically map ``elite'' high-performing solutions across feature dimensions, constructing a rich landscape of solutions~\cite{mouret2015illuminating}. Although effective in avoiding local optima and exposing a broad range of solutions, MAP-Elites can be computationally demanding. Expanding on this foundation, Gravina et al. introduced ``surprise'' as an additional QD criterion, further encouraging exploration of underrepresented areas in the behavioral space~\cite{gravina2018quality}. Newer multi-objective QD extensions like MOME~\cite{pierrot2022multi}, which saves a Pareto front of solutions for each cell in the behavioral space, and Bayesian-optimized methods like BOP-Elites~\cite{kent2024bayesian}, bring QD techniques to more complex and\slash costly fitness evaluation scenarios.

Cully et al.~\cite{cully2017quality} proposed a modular framework for QD algorithms, later enhanced and implemented in the Python library pyribs~\cite{tjanaka2023pyribs}. This framework rests on three core components: behavioral descriptors, the archive, and emitters. The archive is a data structure collecting diverse candidate solutions based on their location in the behavioral descriptor space rather than strictly by performance. Solutions enter the archive if they surpass a novelty threshold, occupy a new part of the behavioral space, or outperform similar solutions in quality. Emitters, which generate or select solutions for the archive, operate with different strategies, prioritizing metrics like quality, novelty, or curiosity. Unlike traditional genetic operators, emitters can represent entire sub-optimization processes aimed at increasing solution quality or exploring new behavioral areas, thus offering a flexible approach to populating the behavioral space with diverse solutions.

The concepts of archive and behavioral descriptors play a central role in both the design and understanding of SETBVE. Behavioral descriptors, in particular, can range from generic metrics (e.g., the number of exceptions thrown by a pair of inputs) to more SUT-specific measures such as the number of classes revealed in a simple classification task. Throughout Section~\ref{sec:approach} we explain how these quality-diversity concepts translate effectively into SETBVE.

\subsubsection{Quality Diversity Optimization in Testing}
QD algorithms have recently gained attention in software testing for their ability to explore diverse and high-performing test scenarios simultaneously. Novelty Search was first applied to test data generation in search-based structural testing by Boussaa et al.\cite{boussaa2015noveltysearch}, who demonstrated its potential for exploring large input spaces. Instead of relying on fitness-based selection, their approach prioritizes test cases with high novelty scores --- those that differ significantly from previously evaluated solutions. 

Building on this idea, Marculescu et al.\cite{marculescu2016using} compared exploration-based algorithms, including Novelty Search and MAP-Elites, with an objective-based approach for testing a clustering algorithm. Their results showed that exploration-focused methods cover a broader portion of the behavior space and produce more diverse solutions, with over 80\% of their outputs not found by the objective-based method, even under limited resources. They concluded that such algorithms are well-suited for investigating high-dimensional spaces, especially when information or computational power is constrained. 

Further reinforcing the benefits of QD in testing, Feldt and Poulding~\cite{feldt2017searching} investigated various methods for generating test inputs that exhibit high diversity with respect to specific features, including techniques inspired by the general principles of Novelty Search and MAP-Elites. Extending these ideas to test suite generation, Xiang et al.~\cite{xiang2023automatedspl} applied the MAP-Elites algorithm to Software Product Lines (SPLs), combining objective functions with a user-defined behavior space. Their approach outperformed traditional single- and multi-objective methods, producing a wide range of effective and diverse test suites that support more informed decision-making.

Beyond traditional software, QD algorithms have also been applied to testing deep learning systems. Riccio and Tonella~\cite{riccio2020model} developed DeepJanus, a tool that generates input pairs with similar features but differing behaviors to map the behavior frontier of a DL system. By combining NSGA-II with Novelty Search, DeepJanus promotes diverse behavior discovery and avoids local optima, helping developers assess system quality and identify inputs that the system fails to handle properly. Building on similar input scenarios, Zohdinasab et al. ~\cite{zohdinasab2021deephyperion} developed DeepHyperion, an open-source test input generator that leverages Illumination Search to produce diverse, high-performing test cases. The approach explores the feature space of DL inputs by mapping interpretable input characteristics to behavioral outcomes, helping developers understand how structural and behavioral features affect system performance. As a result, DeepHyperion provides a human-interpretable view of system quality and helps uncover misbehaving or near-misbehaving cases across a range of feature dimensions. In follow-up work~\cite{zohdinasab2023efficient}, the authors introduced DeepHyperion-CS, an improved version of DeepHyperion for testing DL systems. Instead of selecting inputs randomly, it prioritizes those with higher contribution scores --- inputs more likely to expand feature space coverage. Experiments showed that DeepHyperion-CS outperforms the original tool in both efficiency and effectiveness at uncovering misbehaving inputs.

In summary, QD algorithms have recently shown strong potential in testing, with successful applications in both traditional systems and deep learning models. Their appeal lies in the ability to flexibly define a behavioral space and efficiently explore large input domains, aiming not only to discover diverse solutions but also to ensure high quality within each explored region. Despite the growing interest in QD for testing, to the best of our knowledge, its application to BVE remains unexplored. Given the nature of BVE --- where discovering diverse yet meaningful boundaries is key --- QD offers a promising foundation. In this work, we propose an automated black-box BVE framework that integrates QD algorithms to search for diverse boundary candidates across behavioral space.

%% file: files/approach.tex
\section{SETBVE Approach}
\label{sec:approach}

\begin{figure}
    \centering
    \includegraphics[width=0.9\textwidth]{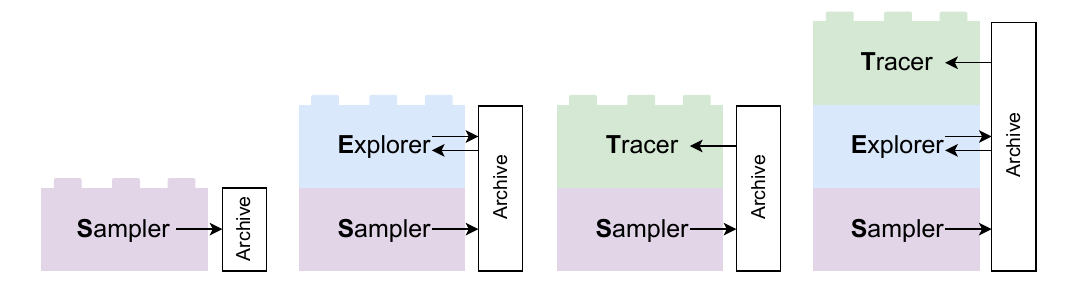}
    \caption{Components of SETBVE, their combinations and interactions with the archive. Arrows indicate solution generation (to archive) and selection (from archive).}
    \label{fig:componentsSETBVE} 
\end{figure}

We propose SETBVE, a framework for automated black-box BVE that incorporates elements of QD approaches. SETBVE aims to identify diverse boundaries and generate multiple example input pairs along each boundary, with the goal of simplifying the boundary summarization phase. While it still uses the program derivative as a metric for evaluating boundariness, similar to AutoBVA, SETBVE introduces a novel approach to boundary search, consisting of three main components --- \textbf{S}ampler, \textbf{E}xplorer, and \textbf{T}racer --- that can be combined in different ways (see Figure \ref{fig:componentsSETBVE}).

The Sampler component generates random solutions and stores them in an archive, which serves both as a record of previously evaluated candidates and as a guide for further exploration. The archive helps ensure that a diverse set of boundary candidates is maintained throughout the search. In our implementation, the archive is structured as a multidimensional grid, where each cell corresponds to a unique combination of values defined by behavioral descriptors. This grid-based structure supports diversity-aware storage and facilitates easier summarization. SETBVE iteratively evaluates input-output pairs, making boundary candidates available either immediately after the search or in real time as the process progresses.

The Explorer component modifies the sampled boundary candidates from the archive and explores the feature space with the aim of populating as many archive cells as possible within a specified time limit. Instead of focusing solely on high program derivative regions, Explorer promotes diversity as well, potentially reducing the risk of overlooking meaningful boundaries with lower derivative values but distinct characteristics. Its search strategy is guided by QD approaches, which consider both feature space diversity and quality. Additionally, our implementation of SETBVE uses Jaccard distance for output distance calculation, because it effectively captures variations in strings compared to simpler metrics.

\begin{figure}
    \centering
    \includegraphics[width=0.9\textwidth]{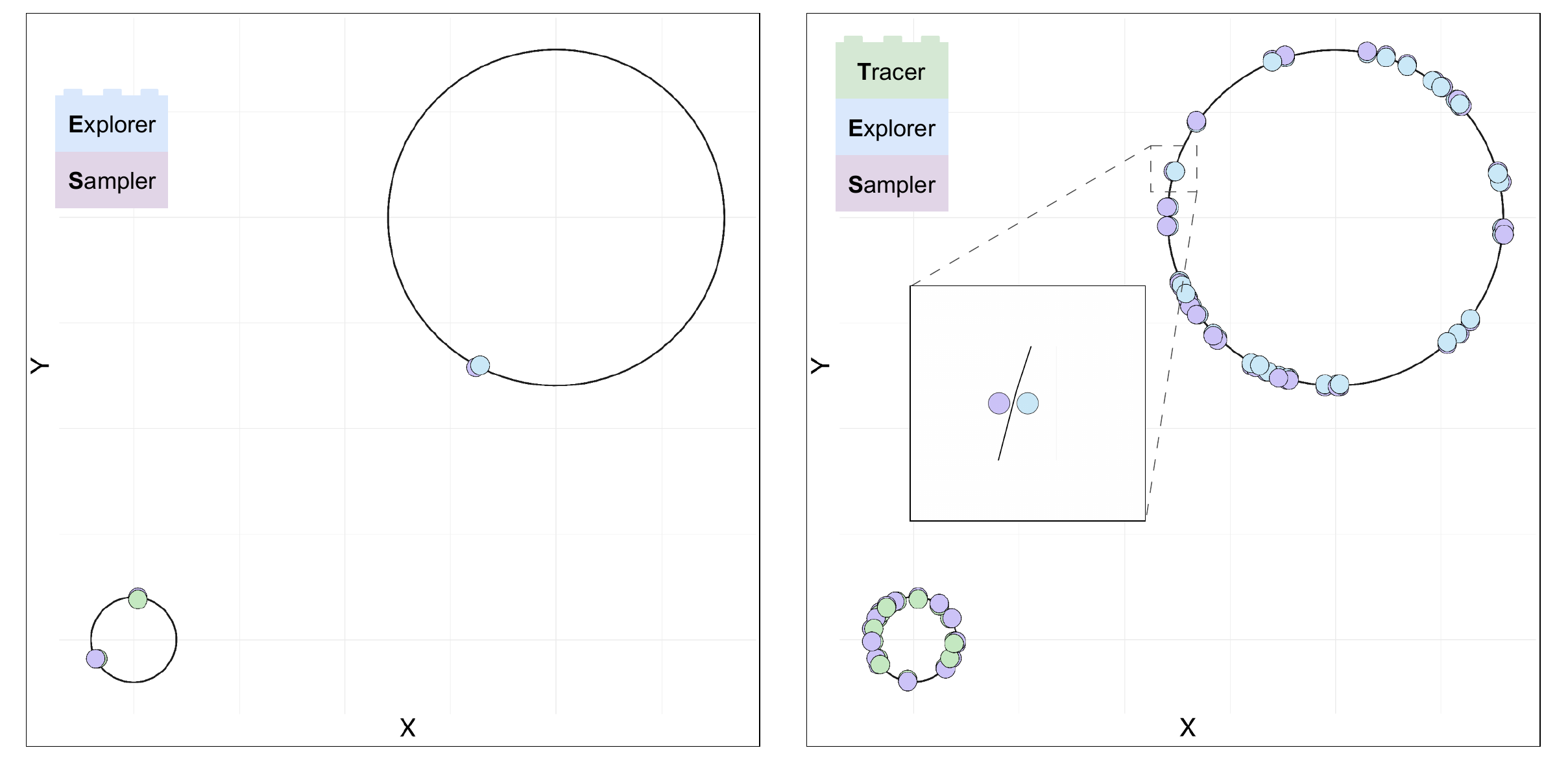}
    \caption{ Illustrative example of boundary refinement. On the left: output from using the Sampler and Explorer. On the right: output after applying the Tracer to the results from the Sampler and Explorer.}
    \label{fig:illustrativeExample} 
\end{figure}

The archive allows only one input pair per cell, making it inherently coarse. The Tracer component refines the search by identifying additional boundary candidates in the vicinity of those already stored in the archive. This enables SETBVE to more thoroughly populate regions near known boundaries, increasing the density of boundary candidates around transition areas. Figure \ref{fig:illustrativeExample} illustrates this refinement where the two circles represent the boundaries. The SUT takes x and y coordinates as input and outputs a string based on the coordinates’ position relative to the circles: ``insideA'', ``insideB'', or ``outsideBoth''. The left side of the figure shows the output of using the Sampler and Explorer, which locate the boundaries of both circles. The Tracer then continues from these points, adding more boundary candidates to the \textit{input space}, and attempting to trace the boundary. The result of this tracing process is displayed on the right side of the figure. Next, we detail those three main building blocks of SETBVE. 

\subsection{Sampler}

Sampler populates the archive with random solutions. We employ a grid-structured archive, where the dimensions are defined by behavioral descriptors, i.e., numerical representations of specific characteristics or features of a solution (in our case, an input-output pair). These descriptors partition the search space into distinct cells, with the framework’s ultimate goal being to discover as many cells as possible (diversity) while ensuring the quality within each discovered cell.

For this version of SETBVE, we propose five behavioral descriptors: three generally applicable across all tested SUTs, and two selective descriptors that vary according to the output type of a SUT. The general descriptors are the number of exceptions, total input length, and input length variance, while the selective descriptors are output abstraction number and output length difference.

\begin{itemize}
    \item \textbf{Number of exceptions:} Since we consider pairs of input values, this descriptor capture three possible regions of the SUT represented in validity groups (see Section~\ref{sec:background}):  Valid-Valid (VV) for input pairs that raise \textit{zero} exceptions, Valid-Error (VE) for pairs raising \textit{one} exception, and Error-Error (EE) when \textit{both} inputs raise exceptions. This descriptor focuses on the output.

    \item \textbf{Output abstraction number:} This descriptor assigns a unique number to pairs of output classes in alphabetical order. For instance, in our experiments with Body Mass Index (BMI) classifications, the combination of outputA ``Normal'' with outputB ``Overweight'' (and vice-versa) are assigned a value, to distinguish between another combination of outputs (e.g., ``Overweight'' and ``Obese''). If an output includes an error, we treat the exception type (e.g., DomainError or ArgumentError) as a class. Note that this descriptor is only applicable to SUTs with categorical outputs.
        
    \item \textbf{Output length difference:} This descriptor captures the difference in output lengths by first converting outputs to strings. Similar to calculating the output abstraction number, if an output contains an error, we extract and use the exception type as the output. The length difference between outputs is then calculated.

    \item \textbf{Total input length:} This descriptor is calculated by converting all input arguments to strings and summing their lengths. It diversifies inputs by their overall length and also serves as a filter after the search when, for instance, there is interest in specific length of input arguments.
    
    \item \textbf{Input length variance:} Like total input length, this descriptor is calculated by firstly converting input arguments to strings and then computing the variance of their lengths.\footnote{We round the variance to the closest integer because our archive uses integer cells.} In combination with total input length, this descriptor enables searches that yield inputs with both uniform and variable lengths. For example, it can capture cases with uniformly short arguments or cases with one long argument among shorter ones.
        
\end{itemize}

In total, we employ four behavioral descriptors as archive dimensions. For SUTs with categorical outputs, we use the following: number of exceptions, output abstraction number, total input length, and input length variance. For other SUTs, we use the following: number of exceptions, output length difference, total input length, and input length variance. This setup provides two dimensions focusing on output characteristics and two focusing on input characteristics.

\begin{figure}
    \centering
    \includegraphics[width=\textwidth]{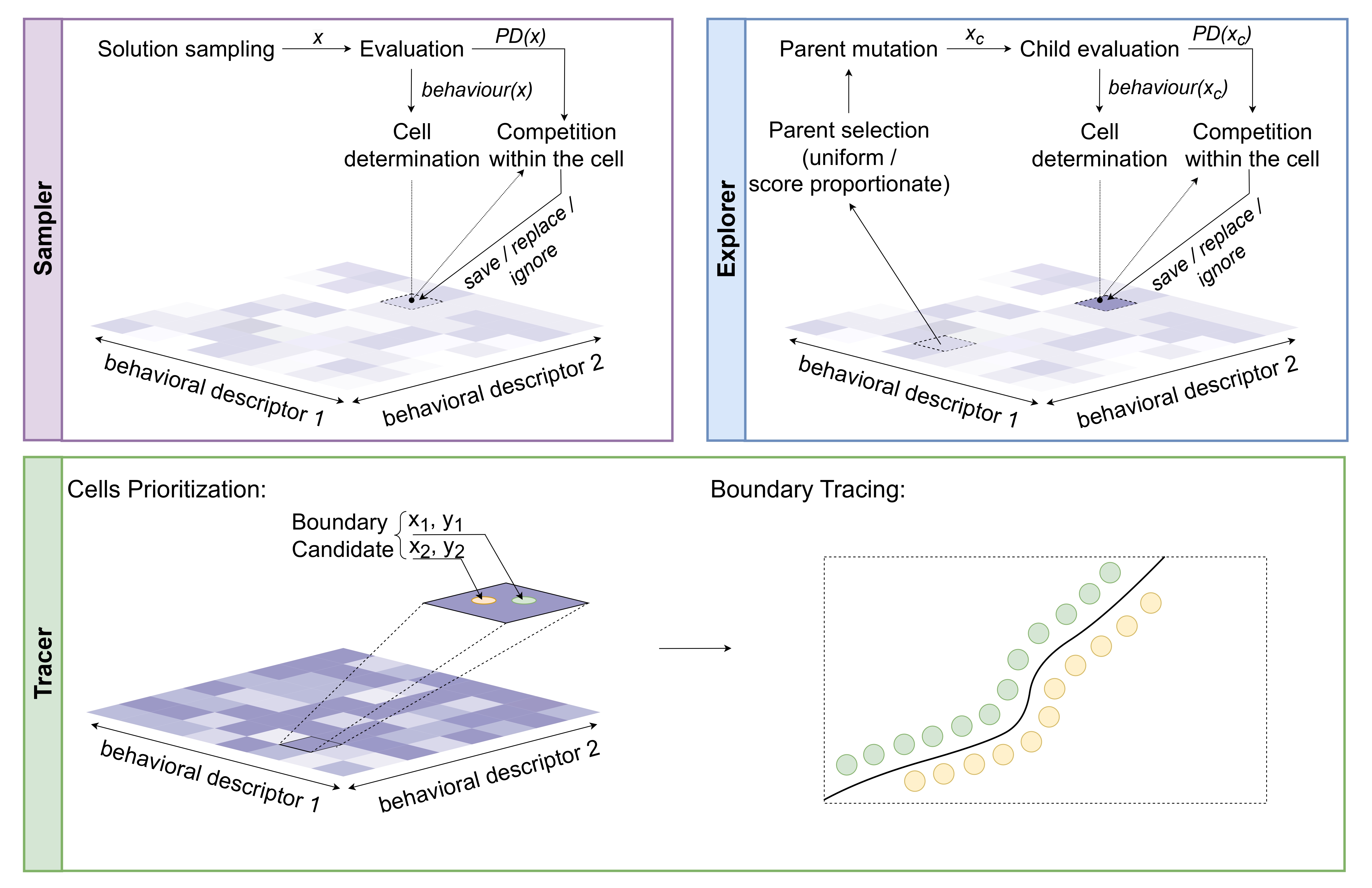}
    \caption{SETBVE framework: illustration of archive sampling, feature space discover process by Explorer (Algorithm \ref{alg:mutation}), and boundary refinement by Tracer (Algorithm \ref{alg:trace}). Darker cells indicate regions with higher program derivative values.}
    \label{fig:approach} 
\end{figure}

The archive population process is illustrated in the top-left section of Figure \ref{fig:approach}. First, the Sampler generates a random solution. For each solution, its behavioral descriptors and program derivative are computed, and the archive is checked to determine whether the corresponding cell is already occupied. If the cell is empty, the solution is stored. If occupied, the solution with the higher program derivative is retained, and the weaker one is discarded. This ensures that each archive cell contains the highest-ranked boundary candidate found for its descriptors.

The Sampler building block is a fundamental component in all combinations of SETBVE elements. When used alone, the archive population process runs throughout the entire time budget. When combined with other blocks, the total time budget is divided between them. In all cases, the process begins with populating the archive (Sampler), and the other blocks are applied afterward.

\subsection{Explorer} 
\label{sec:explorer}
The goal of the Explorer is to \textit{enhance and diversify} the solutions stored in the archive by applying mutations. Diversification is achieved by discovering new cells in the archive by mutating existing boundary candidates, while improvement occurs when a mutated candidate with a higher PD value replaces an existing one in an occupied archive cell.

The process is illustrated in the top right section of Figure ~\ref{fig:approach}, and begins by selecting a parent solution, which is mutated to generate a child solution. A parent solution is selected using one of three methods: (i) uniform random sampling, or score-proportionate selection based on the (ii) solution’s fitness (i.e., program derivative) or (iii) its curiosity score. Uniform random sampling gives each solution an equal chance of selection, whereas in the score-proportionate selection, the program derivative or curiosity scores are used to assign weights to solutions, followed by weighted random sampling.

The curiosity score represents the likelihood that a parent generates an offspring that will be added to the archive ~\cite{cully2017quality}. It is initialized to 0 at the start of the search and updated each time a child is produced. If a child is added to the archive --- either by discovering a new cell or improving an existing solution --- the parent’s curiosity score is increased by 1. Otherwise, if the mutated solution is not added, the curiosity score is decreased by 0.5. 

Once a parent solution is selected, it undergoes mutation outlined in Algorithm ~\ref{alg:mutation}. The mutation starts by bringing the input pair closer together by selecting a random point between the inputs\footnote{We choose a point between $0.25$ and $0.75$ of the distance between the pair. This range was chosen to fairly balance closeness to both parent solutions without biasing toward either extreme.}. A pair to the \texttt{mid\_point} is chosen randomly between the two original inputs (line ~\ref{alg mutation:line:mid_point}). This ensures that the resulting pair is closer than the original. Next, the mutation introduces a randomized shift to one argument of the input pair (line ~\ref{alg mutation:line:random_step}). The shift is a random fraction of the distance between the inputs, chosen at random as positive or negative, creating a relative offset. 

\begin{algorithm}
\caption{Mutation Operator}\label{alg:mutation}
\begin{algorithmic}[1]

\State \textbf{Input:} Boundary candidate $(input1, input2)$ selected from archive 
\State \textbf{Output:} Mutated boundary candidate ${mutated\_bc}$

    \State ${mid\_point} \gets$ Generate random midpoint between ${input1}$ and ${input2}$ \label{alg mutation:line:mid_point}

    \State ${random\_input} \gets$ Randomly select one input from the original pair ${input1}$, ${input2}$ \label{alg mutation:line:random_pair}
    
    \State ${mutated\_bc} \gets$ Form a new boundary candidate by combining ${mid\_point}$ and ${random\_input}$

    \State ${mutated\_bc[idx]} \gets {mutated\_bc[idx]} + step$ (Move a random argument by a random step) \label{alg mutation:line:random_step} 

    \State \Return $mutated\_bc$

\end{algorithmic}
\end{algorithm}

Finally, the behavioral descriptors and the program derivative of the child solution are evaluated to determine whether it should be added to the archive. Similar to the process of populating the archive, the mutated solution is stored if the corresponding cell is empty or if it outperforms the existing solution in terms of program derivative; otherwise, it is discarded. This process repeats iteratively until the allocated time budget is exhausted.

\subsection{Tracer} 
\label{sec:tracer}

Note that the archive stores only the highest-quality solutions for each behaviour defined by the behavioural descriptors, hence becoming intentionally coarse. The goal of the Tracer is to refine the search by closely examining regions around identified boundary candidates to uncover additional solutions in these areas of the \textit{input space}. As shown in the bottom part of Figure ~\ref{fig:approach}, the process consists of two steps: cells prioritization and boundary tracing for the prioritized cells. 

An example of a boundary is illustrated as a curved line in the input space, traced by multiple boundary candidates. These candidates help visualize the boundary’s pattern, potentially offering insights into its shape and structure. The Tracer primarily aims to generate additional solutions near the detected boundary, enabling a more detailed description and analysis for future research.

\subsubsection{Cells Prioritization}

When refining the boundaries, we focus on transitions between equivalence partitions in the VV group and the shift from valid inputs to those causing exceptions (VE). Therefore, we begin by ranking solutions within each validity group based on their program derivative values, ensuring that the most promising boundary candidates are prioritized. From these ranked solutions, a subset of the top candidates\footnote{We use a subset of 100 boundary candidates in this study.} is selected for a focused search in their vicinity within the input space, refining and expanding identified boundaries. To prioritize these cases, candidates are allocated as follows: 50\% from the VV group, 40\% from the VE group, and 10\% from the EE group. If a validity group lacks enough solutions to meet its quota, the shortfall is compensated by selecting candidates from other groups with a surplus.

\subsubsection{Boundary Tracing}

\begin{figure}
    \centering
    \includegraphics[width=0.8\textwidth]{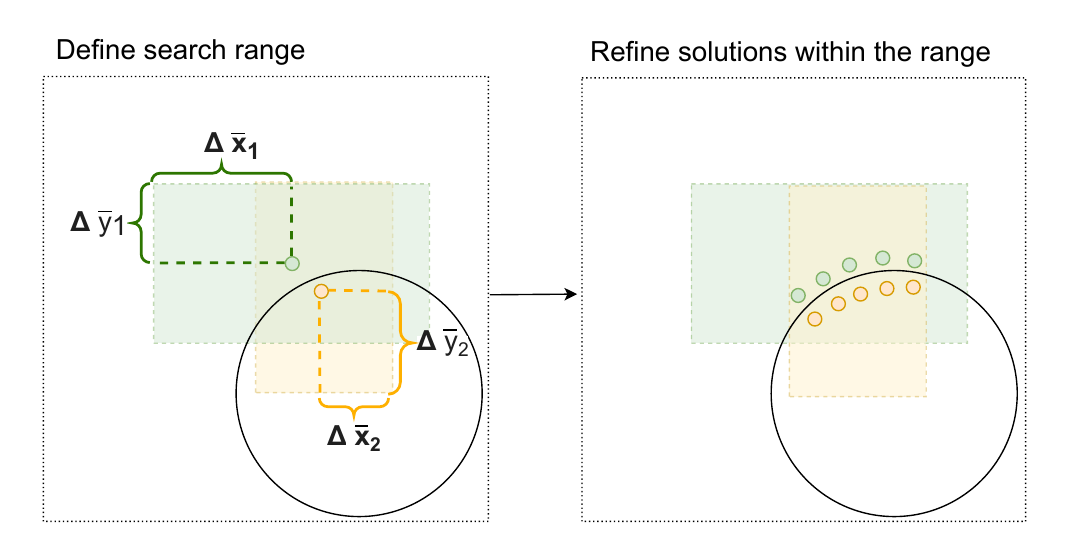}
    \caption{Boundary tracing process.}
    \label{fig:tracingprocess} 
\end{figure}

Figure ~\ref{fig:tracingprocess} illustrates the iterative search process, starting with the first selected boundary candidate. An initial example solution is provided as a starting point. The circle in the figure represents a boundary within a SUT. 

The Tracer component is applied either after the Sampler (when the Explorer is not used) or after the Explorer. Once the Sampler or Explorer has been applied, a list of solutions is produced and sorted in descending order based on their program derivative values. 

The tracing process begins with the first input pair ($x_1$, $y_1$, $x_2$, $y_2$), around which a search region is defined in the \textit{input space}. To determine the search range for each input argument, we compute the differences between 30 consecutive input pairs (sorted by program derivative), starting from the one currently being analyzed. The median of these differences is used as an offset, applied in both positive and negative directions around the selected input pair, thereby defining the search region. This range is illustrated in Figure ~\ref{fig:tracingprocess} as colored rectangles surrounding the input pair, representing the area explored during boundary refinement.

Once the search bounds are established, we conduct a single-objective optimization process to identify boundary candidates that meet the desired criteria within these bounds. The right part of Figure ~\ref{fig:tracingprocess} illustrates the ideal outcome, where boundary candidates are well-spaced and aligned with the boundary. Boundary candidates are input pairs that are close to one another but produce distinctly different outputs (e.g., falling inside and outside the circle). The goal of this search process is to generate 30 boundary candidates inside the search region that balance these objectives.

We define an objective function that balances program derivatives and distances to achieve both the spread between boundary candidates and the closeness within input pairs. Specifically, the objective aims to maximize the total sum of program derivatives across the 30 newly generated boundary candidates while also maximizing the overall distance between them. Since program derivatives are normalized to values between 0 and 1, and distances can vary widely, we scale the sum of the 30 program derivatives by a weight ($W$), representing the maximum possible distance between any two solutions in the search space. We calculate $W$ as the Euclidean distance between the minimum and maximum values within the search bounds ($x1_{\text{min}}$, $y1_{\text{min}}$, $x2_{\text{min}}$, $y2_{\text{min}})$ and ($x1_{\text{max}}$, $y1_{\text{max}}$, $x2_{\text{max}}$, $y2_{\text{max}}$), resulting in Equation \ref{eq: obj_fn_30sols}, our complete objective function.

\begin{equation}
    \text{Objective Function}\ (f_{obj}) = (W \cdot \sum_{i=1}^{30} \text{PD}(BC_i)) + \sum_{\substack{i,j=1 \\ i \neq j}}^{30} \text{Distance}(BC_i, BC_j)
    \label{eq: obj_fn_30sols}
\end{equation}

where:
\begin{itemize}
    \item \( W \) is the weight calculated as the maximum Euclidean distance within the search space: 
    \[ W = \sqrt{(x1_{\text{max}} - x1_{\text{min}})^2 + (y1_{\text{max}} - y1_{\text{min}})^2 + (x2_{\text{max}} - x2_{\text{min}})^2 + (y2_{\text{max}} - y2_{\text{min}})^2} \]
    \item \(\text{PD}(BC_i)\) is the program derivative of boundary candidate \(BC_i\)
    \item \(\text{Distance}(BC_i, BC_j)\) is the Euclidean distance between boundary candidates \(BC_i\) and \(BC_j\)
\end{itemize}

Calculating the entire fitness and the total distance between all solutions in each iteration is computationally expensive. To simplify, we approximate Equation ~\ref{eq: obj_fn_30sols} by calculating the program derivative and distance for just two boundary candidates: a newly generated candidate and a randomly selected one from the current population. This yields Equation \ref{eq: obj_fn_2sols}, which follows the same logic as Equation ~\ref{eq: obj_fn_30sols} but reduces the calculation load from 30 to only two (the newly generated boundary candidate (child) and one selected from the population).

\begin{equation}
    \text{Objective Function}\ (f_{obj}) =( W \cdot (\text{PD}(BC_{\text{child}}) + \text{PD}(BC_{\text{rand}}))) +  \text{Distance}(BC_{\text{child}}, BC_{\text{rand}})
    \label{eq: obj_fn_2sols}
\end{equation}

where:
\begin{itemize}
    \item \( BC_{\text{child}} \) is the newly generated (mutated) boundary candidate
    \item \( BC_{\text{rand}} \) is a randomly selected boundary candidate from the current population, excluding the parent solution
\end{itemize}

The pseudocode for boundary tracing is shown in Algorithm~\ref{alg:trace}. The process iterates over a subset of the \textit{top-$n$} boundary candidates, allocating an equal share of the total tracing time budget ($1/n$) to each. For each candidate, the search starts by initializing a population of 30 solutions within defined bounds. While time remains, a parent solution is randomly selected from the population and mutated using the operator defined in Algorithm~\ref{alg:mutation}.

\begin{algorithm}
\caption{Boundary Tracing}\label{alg:trace}
\begin{algorithmic}[1]

\State \textbf{Input:} Top-$n$ boundary candidates
\State \textbf{Output:} Refined set of boundary candidates

\State $tracing\_populations \gets$ empty list

\For{each boundary candidate in top-$n$ candidates}
    \State Initialize population ($population$) with 30 random boundary candidates

    \While{allocated time for candidate not expired}
        \State $BC_{\text{parent}} \gets$ randomly sample from the $population$
        \State $BC_{\text{child}} \gets$ mutate $S_{\text{parent}}$ using Algorithm~\ref{alg:mutation}
        
        \State $f_{\text{parent}} \gets$ evaluate objective function (Eq.~\ref{eq: obj_fn_2sols}) on $BC_{\text{parent}}$ \label{alg trace:line:obj_parent} 
        \State $f_{\text{child}} \gets$ evaluate objective function (Eq.~\ref{eq: obj_fn_2sols}) on $BC_{\text{child}}$
        
        \If{$f_{\text{child}} > f_{\text{parent}}$}
            \State Replace $BC_{\text{parent}}$ with $BC_{\text{child}}$ in $population$
        \EndIf \label{alg trace:line:end_replace_parent}
    \EndWhile

    \State Append $population$ to $tracing\_populations$
\EndFor

\State \Return $tracing\_populations$

\end{algorithmic}
\end{algorithm}

For each mutation, Equation ~\ref{eq: obj_fn_2sols} is used to calculate the objective function for both the parent and mutated solutions. If the mutated solution yields a higher objective function score, it replaces the parent solution in the population (lines \ref{alg trace:line:obj_parent} -- ~\ref{alg trace:line:end_replace_parent}). Otherwise, the search resumes by selecting a new parent to mutate, with selection occurring uniformly at random. This process continues until the time for the current boundary candidate expires, after which it repeats for the next boundary candidate.

%% file: files/methodology.tex
\section{Empirical Evaluation}
\label{sec:methodology} 

To evaluate the proposed framework’s effectiveness in identifying boundary candidates, we follow Wohlin et al.’s \cite{wohlin2012experimentation} guidelines for software engineering experiments. Our evaluation focuses on comparing different search approaches aiming to identify high-quality boundary candidates while emphasizing their diversity from the perspective of testers or developers. Specifically, we investigate the following research questions:

\begin{itemize}
    \item \textbf{RQ1}: How effective are the search strategies in identifying boundaries in terms of both quality and diversity?
    \item \textbf{RQ2}: How do the search strategies differ in their ability to discover unique regions of the behavioral space?
    \item \textbf{RQ3}: Can the Tracer component further delineate identified boundary regions?
\end{itemize}

For RQ1 and RQ2, we evaluate boundary detection performance across various SETBVE configurations and compare it with AutoBVA --- the only existing automated boundary detection framework. Although AutoBVA is the most relevant baseline, it is not directly comparable due to differences in framework design. To reduce these differences and enable a more meaningful comparison, we take two steps: (1) selecting a subset of relatively high-PD solutions for analysis, and (2) mapping AutoBVA’s solutions to the same archive cell structure used in SETBVE.

To define a this subset, we establish a \textit{boundariness threshold} for each SUT based on all experimental runs, selecting solutions with a program derivative of 1\footnote{The maximum possible PD is 1 due to normalized output distances ranging from 0 to 1}, along with the top 1\% of remaining solutions ranked by PD across all search strategies. Additionally, to enable a direct comparison, we map AutoBVA’s solutions to corresponding archive cells by computing behavioral descriptors for each input-output pair.

RQ1 examines how different variations of SETBVE and AutoBVA perform in identifying boundary candidates, comparing their effectiveness in terms of quality and diversity. We assess: (1) the boundariness of the solutions (i.e., the ability to show change in the SUT's behavior) and (2) archive coverage (i.e., the extent to which the search space is explored, reflecting diversity). To quantify these aspects, we introduce specific measures, detailed in Section \ref{sec:metrics}.

RQ2 focuses on comparing search strategies based on their ability to discover boundary candidates that exhibit distinct and unique behaviors. These behaviors are defined using archive dimensions, which capture input and output characteristics in our experiments. These descriptors can be customized by developers or testers to align with their specific priorities. To address this question, we evaluate how many archive cells ----- containing relatively high-PD solutions --- each search strategy uniquely discovers. To compare strategies from a global perspective, we aggregate results across all runs for each SUT within a fixed time budget allocated to each search strategy.

Lastly, RQ3 focuses on the Tracer component of the SETBVE framework, assessing its ability to refine and expand identified boundary regions. Specifically, we evaluate whether it can discover additional boundary candidates that are close to existing ones and that contribute to a more complete distribution along the boundary. To measure its effectiveness, we visualize boundary regions before and after applying the Tracer, analyzing improvements in coverage and distribution.

\subsection{Quality and Diversity Metrics} \label{sec:metrics}

To enable consistent comparison between different search approaches across various SUTs, we introduce two key metrics that assess both solution performance and behavioral space coverage: \textit{Relative Program Derivative (RPD)} and \textit{Relative Archive Coverage (RAC)}.

\subsubsection{Relative Program Derivative (RPD)} \label{sec:RPD}
RPD quantifies the quality of a boundary candidate by normalizing its PD relative to the \textit{best observed} PD within its \textit{corresponding archive cell} for a given SUT. This normalization addresses the gap between the theoretical maximum ($PD = 1$) and the empirically highest PD observed, which can vary across behavioral domains (i.e., archive cells). Since calculating the exact maximum PD for each cell is infeasible, we approximate it using the highest PD observed across all empirical runs. 

To illustrate this, consider a SUT that takes a single integer input and returns either ``Positive'' for numbers greater than 0 or ``Negative'' for numbers less than 0. For the input pair \texttt{i1 = -1, i2 = 1}, the outputs are \texttt{o1 = "Negative", o2 = "Positive"}, respectively. Since the function only accepts integers, this pair represents the closest input pair resulting in a change from ``Negative'' to ``Positive''. For example, using Jaccard distance based on 2-grams, the output distance is 0.73, and the input distance is 2, yielding a PD of 0.37.

Now, assume we use a grid archive defined by two dimensions: output length difference and total input length. The corresponding cell with coordinates  $(0,3)$ (where 0 represents no output length difference, and 3 is the total input length) has a maximum achievable PD of 0.37. Suppose another boundary candidate is identified with the input pair -2 and 1, resulting in $PD = 0.24$ but assigned to the same archive cell (0,3). In the first case, the $RPD_{(-1,1)} = 0.37/0.37 = 1$, while the $RPD_{(-2,1)} = 0.24/0.37 = 0.65$. Note that, $RPD_{(-1,1)} > RPD_{(-2,1)}$.

The choice of distance functions for calculating the program derivative depends on the input and output types of the SUT. In our experiments, since the inputs are numeric, we use Euclidean distance to measure input differences. For output differences, we employ Jaccard distance based on 2-grams \cite{jaccard1912distribution}, which captures string variations while remaining computationally efficient for short outputs in our experimental SUTs. 

\subsubsection{Relative Archive Coverage (RAC)} \label{sec:RAC}
Similar to RPD, it is infeasible to precisely estimate which archive cells can be covered. RAC quantifies the diversity of solutions by measuring the proportion of the behavioral space covered relative to the empirically observed maximum coverage. The concept and motivation behind RAC align with the notion of \textit{coverage} described in \cite{mouret2015illuminating}, but due to the broad and varied use of the term \textit{coverage} in software testing, we adopt the term RAC to avoid ambiguity.

The need for RAC arises from the fact that, depending on the SUT and the chosen behavioral descriptors, some archive cells may be inherently unpopulatable. For example, consider a grid archive defined by two dimensions: output length difference and total input length, applied to a SUT with a single integer input. While it is possible to populate a cell where the output string length difference is 0 (i.e., two outputs of equal length), the total input string length for a solution (pair of inputs) cannot be less than 2, as each input must have at least one character. As a result, cells with a total input length of 0 or 1 are impossible to populate in the archive.

We normalize archive coverage using the empirically observed maximum to allow for consistent comparisons of coverage across different SUTs and search strategies. Certain combinations of behavioral descriptor values may be unattainable due to the constraints of the SUT. Therefore, we approximate the feasible archive cells by aggregating all discovered archive cells across all runs and search strategies.

\subsection{Description of SUTs} \label{sec:suts}

We evaluate the search approaches on a total of ten SUTs, all operating at the unit level and taking integer inputs, though they differ in input structure and output behavior. Six of these SUTs are taken from Julia’s core library, Base, and each accepts two numeric arguments as input. These include \texttt{cld} (computes the ceiling of the division of two numbers), \texttt{fld} (computes the floor of the division), \texttt{fldmod1} (returns both floor division and modulo as a pair), \texttt{max} (returns the greater of two values), \texttt{power\_by\_squaring} (raises the first argument to the power of the second), and \texttt{tailjoin} (joins the types of elements from a given index to the end). Although the AutoBVA paper evaluated a larger number of Base functions, these six were specifically shortlisted in that work. For these SUTs, we report base performance characteristics using our metrics.

We perform a more detailed evaluation --- both quantitative and qualitative --- on four additional SUTs: \texttt{bytecount}, \texttt{circle}, \texttt{bmi}, and \texttt{date}. Below, we provide the rationale for selecting each SUT, along with details on their inputs and outputs. The code for each SUT was implemented in Julia and is available in our replication package\footnote{\url{https://github.com/aksabina/SETBVE}}.

\begin{itemize}
    \item \texttt{bytecount (bytes: Int)}: Takes an integer representing the number of bytes as input and returns a human-readable string for valid inputs. If the input is out of bounds, the function throws an exception.
    \item \texttt{circle (x: Int, y: Int)}: Receives two integers representing x and y coordinates on a plane, where a circle is centered at the origin. The function determines whether the given coordinates fall inside or outside the circle, returning “in” or “out”, respectively. If the input corresponds to the center of the circle, an exception is thrown.
    \item \texttt{bmi (h: Int, w: Int)}: Takes two integers as input: \texttt{h} represents a person’s height in centimeters, and \texttt{w} represents their weight in kilograms. The function returns a string indicating the BMI category, ranging from underweight to severely obese. If either input is negative, the function throws an exception.
    \item \texttt{date (d: Int, m: Int, y: Int)}: Receives three integers representing day, month, and year. The function returns a string representing the corresponding date in the proleptic Gregorian calendar. If the input forms an invalid date, an exception is thrown, specifying which argument caused the error.
\end{itemize}

Each SUT presents unique testing challenges, differing in input structure and boundary complexity. \texttt{bytecount} focuses on transitions between byte-scale boundaries, where rounding and threshold effects may arise. \texttt{circle} works in a two-dimensional space, where an out-of-range value in either input renders the other irrelevant, defining a finite, well-defined boundary. The \texttt{bmi} SUT increases complexity, as the output depends on the combination of height and weight, with continuous rather than finite boundaries. Lastly, \texttt{date} has three interdependent inputs (day, month, and year), introducing dynamic constraints such as leap years and varying month lengths.

\subsection{Experimental Setup}

We compare the performance of AutoBVA and different configurations of SETBVE by combining its building blocks: Sampler, Explorer, and Tracer. This comparison serves a form of ablation study to better understand the relative importance and contribution of each component in the framework. Table \ref{tab:configurations} summarizes the experimental setup for each search strategy. We implement the search strategies and the experiment instrumentation in Julia\footnote{\url{https://julialang.org/}}. 

\begin{table}
\caption{The different configurations used in search strategies}
\label{tab:configurations}
\centering
\scriptsize
\renewcommand{\arraystretch}{1.2} 
\begin{tabular*}{\textwidth}{@{\extracolsep{\fill}}l c cccc}
\toprule
 & & \multicolumn{4}{c}{\textbf{SETBVE}}\\ 
\cmidrule(lr){3-6}
  \textbf{Parameters} & \textbf{AutoBVA}  &  \textbf{S} & \textbf{SE} & \textbf{ST} & \textbf{SET} \\ 
\midrule 

Sampling strategy & CTS+BU & CTS+BU/\texttt{rand(Int64)} & CTS+BU & CTS+BU & CTS+BU \\ 
Parent selection & --- & --- & curiosity/fitness/uniform & --- & curiosity/fitness/uniform\\ 
Mutation operators & Algorithm 1 in \cite{dobslaw2023automated} & --- & Algorithm \ref{alg:mutation} & Algorithm \ref{alg:mutation}  & Algorithm \ref{alg:mutation} \\ 
Output distance & \texttt{strlendist} & \texttt{jaccard} & \texttt{jaccard} & \texttt{jaccard} & \texttt{jaccard} \\
S,E,T budget distribution & --- & \{100, 0, 0\} \% & \{10, 90, 0\} \% & \{90, 0, 10\} \% & \{10, 80, 10\} \% \\ 
\bottomrule
\end{tabular*}
\begin{tablenotes}
    \item CTS: Compatible Type Sampling
    \item BU: bituniform sampling
    \item S,E,T: Sampler, Explorer, Tracer.
\end{tablenotes}
\end{table}

When evaluating the Sampler component alone, we compare two methods for populating the archive: (1) random uniform sampling of integer values that use 64 bits (\texttt{Int64}) as a baseline, and (2) a combination of CTS with bituniform sampling (Section~\ref{sec:background}), which was identified as the best-performing sampling method in previous work with AutoBVA~\cite{dobslaw2023automated}. When the Sampler is combined with other components, we use CTS with bituniform sampling for creating the initial population. 

The Explorer component applies different parent selection strategies, which include: uniform random selection (randomly choosing parents), fitness-biased selection (favoring high-performing solutions), and curiosity-biased selection (favoring parents that are likely to generate novel or improved offspring). SETBVE mutates solutions using the operator described in Algorithm ~\ref{alg:mutation}, whereas AutoBVA employs two types of search strategies: Local Neighbourhood Search (LNS) and Boundary Crossing Search (BCS), both using increment and decrement operations for numeric inputs. In this work, we use the results of the BCS search for AutoBVA, as it performed better than LNS.  Another key distinction between  AutoBVA and SETBVE is the choice of output distance function to calculate PDs. AutoBVA uses the difference in output length, whereas SETBVE uses the Jaccard distance.

For the results comparability, we reuse the time budgets used in previous experiments with AutoBVA, namely: 30 seconds and 600 seconds. In SETBVE, the budget is distributed among its components. When only the Sampler is used, the entire budget is dedicated to solution sampling. When the Sampler is combined with the Explorer, 10\% of the budget is allocated to initializing the archive, while the Explorer utilizes the remaining 90\% to explore the search space through mutation for boundary identification. In the configuration combining the Sampler and Tracer, 90\% of the budget is allocated to solution sampling, while the Tracer refines the found boundaries with the remaining 10\%.  When all three components are combined, the budget allocation is as follows: 10\% each for the Sampler and Tracer, with the Explorer receiving 80\% because it deserves more time to discover different regions of the behavioral space. We leave the optimization of these budget allocations for future work. In total, we evaluate \textit{nine SETBVE configurations}, derived from four different combinations of building blocks (see Figure~\ref{fig:componentsSETBVE}), two sampling strategies, and three parent selection strategies for each SUT and time budget. 

The experiments are conducted on ten SUTs described in Section~\ref{sec:suts}. Nine of these SUTs were used in the AutoBVA study, and one additional SUT (\texttt{circle}) is introduced in this paper. For AutoBVA results, we use the publicly available data for \texttt{bytecount}, \texttt{bmi}, and \texttt{date}\footnote{\url{https://doi.org/10.5281/zenodo.7677012}}, and we run the AutoBVA framework ourselves for the new \texttt{circle} SUT. In addition, while the results for \texttt{cld}, \texttt{fld}, \texttt{fldmod1}, \texttt{max}, \texttt{power\_by\_squaring}, and \texttt{tailjoin} are available in the AutoBVA replication package, they are limited to 30-second runs. For our empirical evaluation, we re-run AutoBVA for these SUTs with a 600-second time budget. Following the protocol used in the AutoBVA study, each search strategy is executed 20 times to account for pseudorandom variation.

%% file: files/results.tex
\section{Results and analysis} \label{sec:results}
In this section, we analyze and compare how well each method covers the behavioral space, the quality of the boundaries identified, their ability to discover unique behaviors, and the impact of boundary tracing on results. Each research question is discussed in a dedicated subsection, which concludes with a summary of key findings.

\subsection{RQ1 - Quality and Diversity of Boundary Candidates}

In RQ1, we use the Relative Program Derivative (RPD) and Relative Archive Coverage (RAC) to compare, respectively, the quality and diversity of boundary candidates identified by different search strategies across the tested SUTs. We evaluate each strategy using two time budgets: 30 seconds and 600 seconds. The results presented in Tables \ref{tab:qdbytecount} -- \ref{tab:qddate} are sorted by RAC at 600 seconds because our primary focus is on the diversity of boundary candidates and exploration of the search space.

All reported values represent the mean and standard deviation from 20 experimental runs. Note that RPD is normalized to a range from 0 to 1, while RAC expresses the percentage of empirically observed archive cells with relatively high PD that are covered. Consequently, a RAC value of 100\% indicates that all such archive cells identified across all runs and search strategies have been discovered.

\subsubsection{Bytecount}

\begin{table}[h]
    \centering
    \scriptsize 
    \caption{Descriptive statistics ($\mu$ ± $\sigma$) for quality (RPD) and diversity (RAC) for the \textbf{\texttt{bytecount}} SUT. The top three results for each metric, including ties, are highlighted in bold. CTS: Compatible Type Sampling, BU: bituniform sampling.}
    \renewcommand{\arraystretch}{1.2}
    \setlength{\tabcolsep}{8pt}   
    \begin{tabular}{llccc|cc|cc}
        \toprule
        & \multicolumn{1}{c}{} & \multicolumn{3}{c|}{\textbf{Configurations}} & \multicolumn{2}{c|}{\textbf{30 seconds}} & \multicolumn{2}{c}{\textbf{600 seconds}} \\
        \cmidrule(lr){3-5} \cmidrule(lr){6-7} \cmidrule(lr){8-9}
         & \textbf{Framework} & \textbf{Sampler} & \textbf{Explorer} & \textbf{Tracer} & \textbf{RPD} & \textbf{RAC, \%} & \textbf{RPD} & \textbf{RAC, \%} \\
        \midrule
        1 & SETBVE & CTS+BU & Curiosity & + & \textbf{0.996} ± 0.003  & \textbf{95.22} ± 0.95 & \textbf{0.997} ± 0.002 & \textbf{100.0} ± 0.0 \\
        \rowcolor{lightgray} 
        2 & SETBVE & CTS+BU & Uniform & --- & 0.977 ± 0.01 & \textbf{94.73} ± 1.2 & \textbf{0.999} ± 0.002 & \textbf{99.95} ± 0.24 \\
        3 & SETBVE & CTS+BU & Uniform & + & 0.97 ± 0.01 & \textbf{94.68} ± 1.33 & \textbf{0.997} ± 0.002 & \textbf{99.95} ± 0.24 \\
        \rowcolor{lightgray}
        4 & SETBVE & CTS+BU & Curiosity & --- & \textbf{0.995} ± 0.002 & \textbf{95.48} ± 1.29 & \textbf{0.999} ± 0.002 & \textbf{99.84} ± 0.53 \\
        5 & SETBVE & CTS+BU & --- & --- & 0.98 ± 0.02 & 70.48 ± 3.53 & 0.98 ± 0.01 & 91.13 ± 1.35\\
        \rowcolor{lightgray}
        6 & SETBVE & CTS+BU & --- & + & 0.96 ± 0.01 & 72.69 ± 2.93 & 0.97 ± 0.01 & 90.0 ± 1.26 \\
        7 & SETBVE & CTS+BU & Fitness & + & 0.975 ± 0.01 & 58.23 ± 1.68 & 0.98 ± 0.01 & 81.88 ± 2.35 \\
        \rowcolor{lightgray}
        8 & SETBVE & CTS+BU & Fitness & --- & 0.976 ± 0.01 & 47.37 ± 1.89 & 0.983 ± 0.01 & 79.57 ± 2.15 \\
        9 & AutoBVA & --- & --- & --- & \textbf{1.0} ± 0.0 & 56.67 ± 0.51 & \textbf{1.0} ± 0.0 & 57.9 ± 0.39\\
        \rowcolor{lightgray} 
        10 & SETBVE & Random  & --- & --- & 0.0 ± 0.0 & 8.28 ± 0.93 & 0.0 ± 0.0 & 10.32 ± 0.81 \\
        \bottomrule
    \end{tabular}
    \label{tab:qdbytecount}
\end{table}

Table \ref{tab:qdbytecount} shows that AutoBVA quickly achieves maximum quality (RPD = 1.0) at 30 seconds, yet its diversity remains relatively low (around 58\%). Random sampling performs poorly, yielding a RPD of zero across both time budgets and low RAC values (8\% at 30s and 10\% at 600s), illustrating its ineffectiveness. 

Similar to the findings in previous studies with AutoBVA, introducing CTS with bituniform sampling enhances the performance, especially when combined with uniform or curiosity-based exploration. These two Explorer options consistently deliver near-perfect quality (RPD $\approx 0.97-0.99$) while also attaining high coverage of candidates (above 94\% at 30s and up to 100\% at 600s). In contrast, fitness-based exploration (rows 7 and 8) is notably less diverse, particularly at 30s (RAC $\approx 47\%-58\%$), though it increases up to 82\% at 600s. Moreover, when using bituniform sampling even without the Explorer and Tracer components, RPD remains high (0.98), and RAC exceeds that of AutoBVA (70\% at 30s and 91\% at 600s). 

The impact of adding Tracer for curiosity-driven and uniform random exploration is minimal in terms of both quality and diversity. In contrast, for fitness-based exploration, the Tracer has a more noticeable impact on RAC, as it improves archive coverage at both time budgets (see rows 7 and 8).

\subsubsection{Circle}
\begin{table}[h]
    \centering
    \scriptsize 
    \caption{Descriptive statistics ($\mu$ ± $\sigma$) for quality (RPD) and diversity (RAC) across search strategies and time budgets for the \textbf{\texttt{circle}} SUT. The top three results for each metric are highlighted in bold. CTS: Compatible Type Sampling, BU: bituniform sampling.}
    \renewcommand{\arraystretch}{1.2}
    \setlength{\tabcolsep}{8pt}   
    \begin{tabular}{llccc|rr|rr}
        \toprule
        & \multicolumn{1}{c}{} & \multicolumn{3}{c|}{\textbf{Configurations}} & \multicolumn{2}{c|}{\textbf{30 seconds}} & \multicolumn{2}{c}{\textbf{600 seconds}} \\
        \cmidrule(lr){3-5} \cmidrule(lr){6-7} \cmidrule(lr){8-9}
         & \textbf{Framework} & \textbf{Sampler} & \textbf{Explorer} & \textbf{Tracer} & \textbf{RPD} & \textbf{RAC, \%} & \textbf{RPD} & \textbf{RAC, \%} \\
        \midrule
        1 & SETBVE & CTS+BU & --- & --- & \textbf{0.7} ± 0.01 & \textbf{52.28} ± 0.46 & 0.84 ± 0.01 & \textbf{72.27} ± 0.88\\
        \rowcolor{lightgray} 
        2 & SETBVE & CTS+BU & --- & + & \textbf{0.71} ± 0.01 & \textbf{51.52} ± 1.04 & \textbf{0.86} ± 0.01 & \textbf{71.45} ± 0.99\\
        3 & SETBVE & CTS+BU & Uniform & --- & 0.55 ± 0.02 & \textbf{38.94}	± 1.08 & 0.84 ± 0.01 & \textbf{57.97}	± 0.78 \\
        \rowcolor{lightgray}
        4 & SETBVE & CTS+BU & Uniform & + & 0.57 ± 0.02 & 38.74 ± 1.02 & \textbf{0.85} ± 0.01 & 57.92 ± 0.94\\  
        5 & SETBVE & CTS+BU & Curiosity & --- & 0.5 ± 0.02  & 37.24	± 0.86 & 0.76 ± 0.02 & 57.74	± 0.97\\
        \rowcolor{lightgray} 
        6 & SETBVE & CTS+BU & Fitness & --- & 0.56 ± 0.02 & 37.03 ± 1.08 & 0.80 ± 0.01 & 57.68 ± 0.71\\
        7 & SETBVE & CTS+BU & Fitness & + & 0.55 ± 0.021 & 37.39 ± 1.06 & 0.8 ± 0.01 & 57.41	± 0.82\\
        \rowcolor{lightgray}
        8 & SETBVE & CTS+BU & Curiosity & + & 0.53 ± 0.02 & 37.43	± 0.97 & 0.77 ± 0.01 & 57.31	± 0.86 \\
        9 & AutoBVA & --- & --- & --- & \textbf{0.94} ± 0.01 & 31.57 ± 0.71 & \textbf{0.95} ± 0.01 & 35.65	± 0.71\\
        \rowcolor{lightgray} 
        10 & SETBVE & Random  & --- & --- & 0.0 ± 0.0 & 2.15	± 0.11 & 0.0 ± 0.0 & 2.63	± 0.1 \\
        
        \bottomrule
    \end{tabular}
    \label{tab:qdcircle}
\end{table}

The results in Table~\ref{tab:qdcircle} show that AutoBVA consistently achieves the highest RPD scores (above 0.94) at both time budgets. However, this comes at the cost of behavioral diversity, with RAC values remaining relatively low --- around 32\% at 30 seconds and 36\% at 600 seconds. In contrast, SETBVE configurations achieve much higher RAC, even without the Explorer or Tracer. For example, the Sampler-only setup (row 1) yields a moderate RPD (0.7 at 30s, 0.84 at 600s), but RAC improves over time, increasing from 52\% to 72\%.

Explorer configurations result in lower RPD scores at 30 seconds (ranging from 0.5 to 0.57) and RAC values clustered around 37–39\%. After 600 seconds, RPD increases across all Explorer variants, with uniform random selection reaching the highest (0.85), followed by fitness- and curiosity-based strategies (0.8 and 0.77, respectively). However, RAC values across all Explorer variants converge in the 57–58\% range. 

Enabling the Tracer yields small improvements in RAC compared with tracer-less settings (e.g., 51.52\% vs. 52.28\% for no Explorer at 30s), while leaving RPD relatively unchanged (rows 1 and 2).

\subsubsection{BMI}
\begin{table}[h]
    \centering
    \scriptsize 
    \caption{Descriptive statistics ($\mu$ ± $\sigma$) for quality (RPD) and diversity (RAC) across search strategies and time budgets for the \textbf{\texttt{bmi}} SUT. The top three results for each metric are highlighted in bold. CTS: Compatible Type Sampling, BU: bituniform sampling.}
    \renewcommand{\arraystretch}{1.2}
    \setlength{\tabcolsep}{8pt}   
    \begin{tabular}{llccc|cc|cc}
        \toprule
        & \multicolumn{1}{c}{} & \multicolumn{3}{c|}{\textbf{Configurations}} & \multicolumn{2}{c|}{\textbf{30 seconds}} & \multicolumn{2}{c}{\textbf{600 seconds}} \\
        \cmidrule(lr){3-5} \cmidrule(lr){6-7} \cmidrule(lr){8-9}
         & \textbf{Framework} & \textbf{Sampler} & \textbf{Explorer} & \textbf{Tracer} & \textbf{RPD} & \textbf{RAC, \%} & \textbf{RPD} & \textbf{RAC, \%} \\
        \midrule
        1 & SETBVE & CTS+BU & Uniform & --- & 0.31 ± 0.02 & 41.15 ± 1.07 & 0.47 ± 0.01 & \textbf{80.2} ± 1.184 \\
        \rowcolor{lightgray} 
        2 & SETBVE & CTS+BU & Uniform & + & 0.35 ± 0.02 & \textbf{44.12} ± 1.25 & 0.48 ± 0.01 & \textbf{79.43} ± 1.01 \\
        3 & SETBVE & CTS+BU & --- & + & 0.4 ± 0.01 & \textbf{48.29} ± 0.64 & 0.54 ± 0.01 & \textbf{77.92} ± 0.81 \\
        \rowcolor{lightgray}
        4 & SETBVE & CTS+BU & --- & --- & 0.38 ± 0.01 & \textbf{48.38} ± 0.86 & 0.51 ± 0.01 & 77.27 ± 1.31 \\
        5 & SETBVE & CTS+BU & Curiosity & + & 0.32 ± 0.02 & 37.47 ± 0.96 & 0.45 ± 0.01 & 60.75 ± 0.88 \\
        \rowcolor{lightgray}
        6 & SETBVE & CTS+BU & Fitness & --- & \textbf{0.5} ± 0.01 & 30.56 ± 1.39 & \textbf{0.56} ± 0.01 & 60.1 ± 0.93\\
        7 & SETBVE & CTS+BU & Fitness & + & \textbf{0.52} ± 0.02 &32.23 ± 1.45 & \textbf{0.56} ± 0.01 & 59.97 ± 0.73 \\
        \rowcolor{lightgray}
        8 & SETBVE & CTS+BU & Curiosity & --- & 0.3 ± 0.02 & 31.4 ± 1.4 & 0.4 ± 0.01 & 59.59 ± 1.27\\
        9 & AutoBVA & --- & --- & --- & \textbf{1.0} ±	0.0 &1.95 ± 0.2 & \textbf{1.0} ± 0.0&3.44 ± 0.11 \\
        \rowcolor{lightgray} 
        10 & SETBVE & Random  & --- & --- & 0.0 ± 0.0 &0.0 ±	0.0& 0.0 ± 0.0 & 0.0 ±	0.0\\
        \bottomrule
    \end{tabular}
    \label{tab:qdbmi}
\end{table}
The results, as detailed in Table \ref{tab:qdbmi}, show that AutoBVA has the highest RPD score of 1.0 at both 30 and 600 seconds, but its diversity remains exceptionally low (RAC $\approx 2\%-3\%$), highlighting its focus on quality at the expense of diversity.

In contrast, SETBVE with bituniform sampling exhibits a trade-off between RPD and RAC depending on the chosen Explorer and Tracer configurations. The highest diversity levels are observed with no Explorer and a Tracer enabled (48\% at 30s, 78\% at 600s), suggesting that this configuration effectively enhances diversity while maintaining a relatively high RPD (0.4 at 30s, 0.54 at 600s).

Among the Explorer strategies, fitness-based exploration produces the highest RPD (0.5 at 30s, 0.56 at 600s) but sacrifices diversity (31\% and 60\%, respectively). When the Tracer is enabled, RPD increases slightly (0.52 at 30s), but RAC remains low. However, curiosity-based exploration leads to lower quality values (RPD $\approx 0.3-0.4$), with RAC ranging between 31\% (at 30s) and increasing to 60\% (at 600s). Uniform random exploration provides a more balanced approach, achieving reasonable RPD values (RPD $\approx 0.3-0.4$) while maintaining one of the highest RAC scores at 600 seconds (79\%–80\%).

\subsubsection{Date}
\begin{table}[h]
    \centering
    \scriptsize 
    \caption{Descriptive statistics ($\mu$ ± $\sigma$) for quality (RPD) and diversity (RAC) across search strategies and time budgets for the \textbf{\texttt{date}} SUT. The top three results for each metric are highlighted in bold. CTS: Compatible Type Sampling, BU: bituniform sampling.}
    \renewcommand{\arraystretch}{1.2}
    \setlength{\tabcolsep}{8pt}   
    \begin{tabular}{llccc|cc|cc}
        \toprule
        & \multicolumn{1}{c}{} & \multicolumn{3}{c|}{\textbf{Configurations}} & \multicolumn{2}{c|}{\textbf{30 seconds}} & \multicolumn{2}{c}{\textbf{600 seconds}} \\
        \cmidrule(lr){3-5} \cmidrule(lr){6-7} \cmidrule(lr){8-9}
         & \textbf{Framework} & \textbf{Sampler} & \textbf{Explorer} & \textbf{Tracer} & \textbf{RPD} & \textbf{RAC, \%} & \textbf{RPD} & \textbf{RAC, \%} \\
        \midrule
        1 & SETBVE & CTS+BU & --- & --- & 0.09 ± 0.004 & \textbf{41.37} ± 0.45 & \textbf{0.24} ± 0.01 & \textbf{79.74} ± 0.87\\
        \rowcolor{lightgray}
        2 & SETBVE & CTS+BU & --- & + & 0.11 ± 0.01 & \textbf{41.78} ± 0.7 & \textbf{0.27} ± 0.01 & \textbf{79.5}	± 0.57\\
        3 & SETBVE & CTS+BU & Uniform & + & 0.1 ± 0.01 & \textbf{35.95} ± 1.1 & 0.21 ± 0.01 &\textbf{74.49} ± 0.37 \\
        \rowcolor{lightgray}
        4 & SETBVE & CTS+BU & Uniform & --- & 0.07 ± 0.01 & 33.09 ± 1.38 & 0.18 ± 0.01 & 73.79 ± 0.73 \\
        5 & SETBVE & CTS+BU & Curiosity & + & 0.08 ± 0.02 & 28.63 ± 1.64 & 0.16 ± 0.01 & 60.08 ± 0.56\\
        \rowcolor{lightgray} 
        6 & SETBVE & CTS+BU & Curiosity & --- & 0.05 ± 0.01 & 21.42 ± 2.46 & 0.12 ± 0.01 & 57.45	± 0.99\\
        7 & SETBVE & CTS+BU & Fitness & + & \textbf{0.19} ± 0.04 & 15.44 ± 1.33 & 0.2 ± 0.01 & 55.28 ± 0.47 \\
        \rowcolor{lightgray} 
        8 & SETBVE & CTS+BU & Fitness & --- & \textbf{0.21} ± 0.03 & 12.69 ± 1.05 & 0.19±0.009 & 52.28	±0.736\\
        9 & AutoBVA & --- & --- & --- & \textbf{0.91} ± 0.02 & 0.53 ± 0.12 & \textbf{0.96} ±	0.002  & 5.47 ± 0.17 \\
        \rowcolor{lightgray}
        10 & SETBVE & Random  & --- & --- & 0.0 ± 0.0 & 0.0 ± 0.0 & 0.0 ± 0.0 & 0.0 ± 0.0\\        
        \bottomrule
    \end{tabular}
    \label{tab:qddate}
\end{table}

Table \ref{tab:qddate} reveals a clear trade-off between quality and diversity for the \texttt{date} SUT. AutoBVA achieves the highest RPD (0.91 at 30s, 0.96 at 600s), but at the cost of extremely low diversity (e.g., only 5.47\% RAC at 600s). In contrast, configurations under SETBVE with bituniform sampling exhibit far higher diversity ($RAC > 50\%$ at 600s) but attain modest RPD scores ($RPD < 0.3$ in all instances).

Among the SETBVE strategies, those without an Explorer consistently yield the highest diversity (over 40\% at 30s and nearly 80\% by 600s), showing that omitting the Explorer can be advantageous for this specific SUT in producing diverse boundary candidates. On the other hand, fitness-based exploration reaches higher RPD values (up to 0.21 at 30s), yet its diversity lags behind (12\% at 30s, and 55\% at 600s). Uniform exploration achieves moderate RPD (starting low at 0.07 and then increasing to 0.21 at 600s) and relatively high diversity at 600s (RAC $= 74\%$). Curiosity-based exploration performs similarly or slightly lower in both RPD and RAC relative to the uniform random parent selection.

Enabling the Tracer generally leads to a marginal increase in RPD and RAC within each exploration category. Finally, random sampling fails to make any progress, yielding zero values in both coverage and diversity. 

\subsubsection{SUTs from Julia Base}

Table~\ref{tab: juliabase_rac_rpd} presents RAC and RPD values averaged over 20 runs for a set of Julia Base functions, comparing AutoBVA with two SETBVE configurations: SE-Uniform (Sampler and Explorer) and SET-Uniform (Sampler, Explorer, and Tracer). In this experiment, we evaluate SETBVE using the uniform parent selection strategy for the Explorer. This choice is based on prior observations from our analysis of four SUTs (\texttt{bytecount}, \texttt{circle}, \texttt{bmi}, and \texttt{date}), where uniform random parent selection delivered competitive results in most cases.

\begin{table}
\centering
\scriptsize
\caption{RPD and RAC ($\mu$ ± $\sigma$) for SUTs from Julia Base. SE-Uniform refers to the SETBVE configuration using a Sampler (CTS with bituniform sampling) and Explorer (uniform random parent selection). SET-Uniform adds a Tracer to this configuration.}
\begin{tabular*}{\textwidth}{@{\extracolsep{\fill}}l l cc cc cc}
\toprule
& & \multicolumn{2}{c}{\textbf{AutoBVA}} &
\multicolumn{2}{c}{\textbf{SE - Uniform}} &
\multicolumn{2}{c}{\textbf{SET - Uniform}} \\
\cmidrule(lr){3-4} \cmidrule(lr){5-6} \cmidrule(lr){7-8}
\textbf{SUT} & \textbf{Time} & \textbf{RPD} & \textbf{RAC, \%} &
\textbf{RPD} & \textbf{RAC, \%} & \textbf{RPD} & \textbf{RAC, \%} \\
\midrule
\noalign{\vskip 3pt}
\multirow{2}{*}{\texttt{cld}} & 30 sec & 0.99 ± 0.002 & 2.4 ± 0.06 & 0.23 ± 0.006 & 44.76 ± 0.53 & 0.24 ± 0.005 & 44.35 ± 0.39 \\
                         & 600 sec & 1.0 ± 0.0 & 1.48 ± 0.05 & 0.39 ± 0.004 & 79.53 ± 0.46 & 0.39 ± 0.008 & 79.06 ± 0.48 \\
\noalign{\vskip 6pt}
\cdashline{1-8}[0.8pt/2pt]
\noalign{\vskip 6pt}
\multirow{2}{*}{\texttt{fld}} & 30 sec & 0.99 ± 0.002 & 2.2 ± 0.06  & 0.22 ± 0.01  & 46.9 ± 0.52  & 0.24 ± 0.01 & 46.43 ± 0.61 \\
                         & 600 sec & 1.0 ± 0.0 & 1.39 ± 0.04 & 0.4 ± 0.008 & 83.91 ± 0.45 & 0.4 ± 0.01 & 83.42 ± 0.48 \\
\noalign{\vskip 6pt}
\cdashline{1-8}[0.8pt/2pt]
\noalign{\vskip 6pt}
\multirow{2}{*}{\texttt{fldmod1}} & 30 sec & 0.94 ± 0.01 & 2.57 ± 0.09 & 0.19 ± 0.004 & 35.49 ± 0.47 & 0.2 ± 0.01 & 34.63 ± 0.88 \\
                         & 600 sec & 0.94 ± 0.005 & 2.49 ± 0.07 & 0.36 ± 0.004 & 78.21 ± 0.22 & 0.36 ± 0.004 & 77.75 ± 0.28 \\
\noalign{\vskip 6pt}
\cdashline{1-8}[0.8pt/2pt]
\noalign{\vskip 6pt}
\multirow{2}{*}{\texttt{max}} & 30 sec & 1.0 ± 0.0 & 1.67 ± 0.06  & 0.28 ± 0.003 & 42.4 ± 0.54 & 0.28 ± 0.004 & 41.98 ± 0.6 \\
                         & 600 sec & 1.0 ± 0.0 & 1.11 ± 0.04 & 0.44 ± 0.005 & 81.65 ± 0.39 & 0.44 ± 0.005 & 81.2 ± 0.39  \\

\noalign{\vskip 6pt}
\cdashline{1-8}[0.8pt/2pt]
\noalign{\vskip 6pt}
\multirow{2}{*}{\texttt{power\_by\_squaring}} & 30 sec & 0.94 ± 0.005 & 2.85 ± 0.07  & 0.2 ± 0.01 & 24.88 ± 0.39 & 0.26 ± 0.01 & 25.12 ± 0.64\\
                         & 600 sec & 0.94 ± 0.005 & 2.58 ± 0.07 & 0.33 ± 0.004 & 60.6 ± 0.63 & 0.33 ± 0.004 & 59.96 ± 0.66 \\ 
                         
\noalign{\vskip 6pt}
\cdashline{1-8}[0.8pt/2pt]
\noalign{\vskip 6pt}

\multirow{2}{*}{\texttt{tailjoin}} & 30 sec & 0.96 ± 0.004 & 1.85 ± 0.04 & 0.25 ± 0.006 & 28.7 ± 0.35 & 0.26 ± 0.01 & 28.03 ± 0.91 \\
                         & 600 sec & 0.96 ± 0.004 & 1.62 ± 0.05 & 0.41 ± 0.005 & 66.19 ± 0.32 & 0.4 ± 0.005 & 65.67 ± 0.42 \\
\noalign{\vskip 3pt}
\bottomrule
\end{tabular*}
\label{tab: juliabase_rac_rpd}
\end{table}

The results show a consistent trend across all SUTs: SETBVE configurations achieve higher diversity (RAC) compared to AutoBVA, indicating their effectiveness in exploring a wider behavioral space. In contrast, AutoBVA consistently attains nearly maximal RPD values, reflecting its emphasis on optimizing individual boundary candidates. This optimization, however, results in lower RAC values, which remain limited even after extended execution (600 seconds). SETBVE configurations improve both RAC and RPD as the runtime increases; for example, RAC values typically grow from approximately 25 -- 45\% at 30 seconds to about 60 -- 80\% at 600 seconds, while RPD rises from around 0.2 to approximately 0.4 over the same period for most SUTs. Including the Tracer component has minimal impact on both RAC and RPD metrics.

\subsubsection{General Trends}

\begin{figure}
    \centering
    \includegraphics[width=\textwidth]{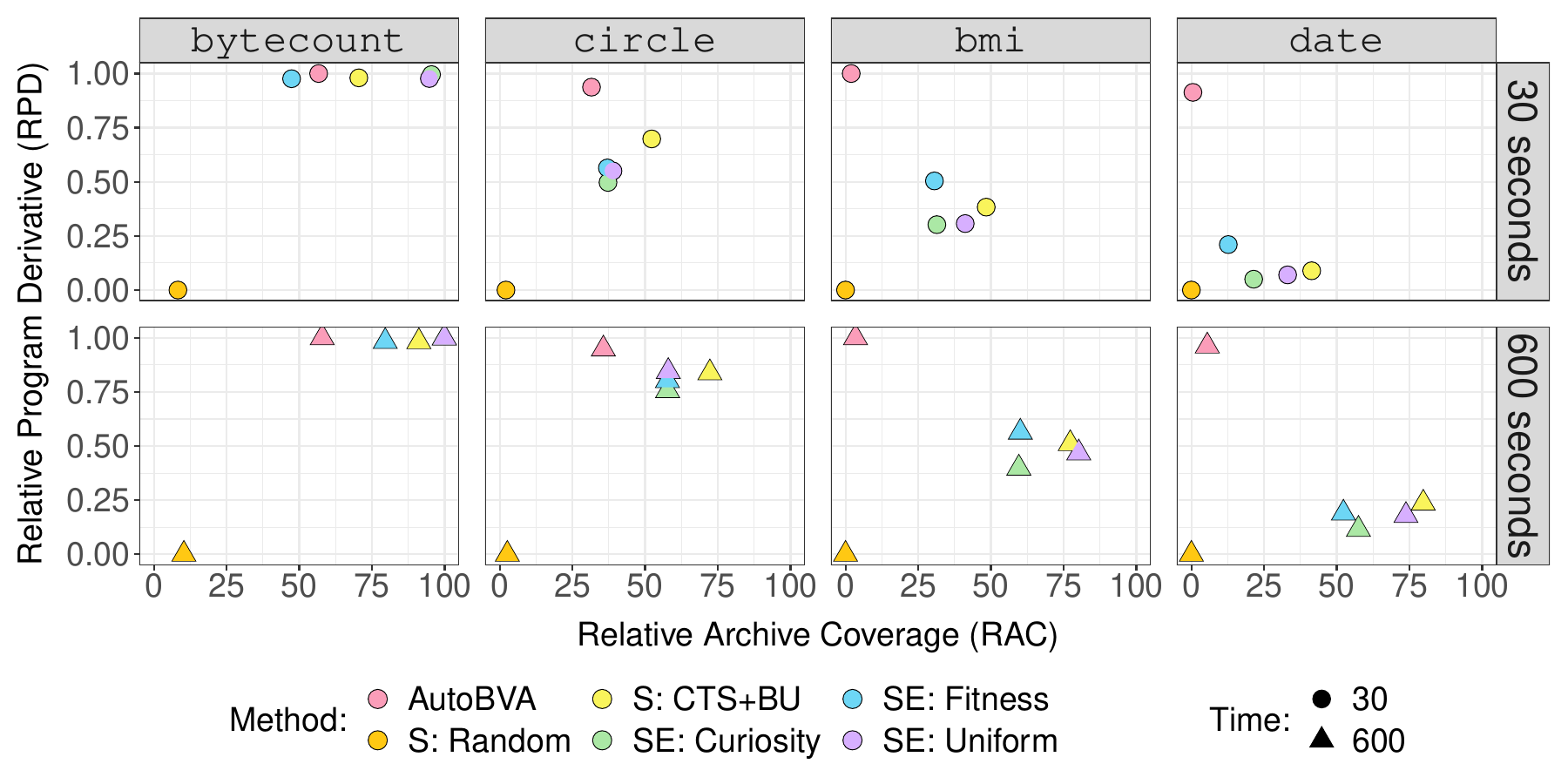}
    \caption{RPD and RAC across methods for \texttt{bytecount}, \texttt{circle}, \texttt{bmi} and \texttt{date}. S: Sampler, E: Explorer, CTS: Compatible Type Sampling, BU: bituniform sampling.}
    \label{fig:RPD_vs_rac_plot} 
\end{figure}

Figure \ref{fig:RPD_vs_rac_plot} illustrates the relationship between RPD and RAC for \texttt{bytecount}, \texttt{circle}, \texttt{bmi}, \texttt{date} SUTs. For detailed numeric values, refer to Tables \ref{tab:qdbytecount}–\ref{tab:qddate}. Methods utilizing the Tracer component are omitted from the figure because Tracer has only a modest impact on RDP and RAC in most tested scenarios. 

A clear pattern emerges when comparing the results at 30 seconds and 600 seconds. Increasing the time budget improves diversity for most methods, as seen in the general shift toward higher RAC at 600 seconds. SETBVE configurations with the Sampler using CTS with bituniform sampling, as well as the Sampler combined with the Explorer across all parent selection strategies, show considerable increases in diversity, indicating that a longer execution allows for broader exploration of the search space. However, AutoBVA and Random sampling do not show much improvement in diversity over time, remaining close to their initial positions. This suggests that these methods reach their maximum performance early on and do not benefit as much from extended execution.  Considering that the RAC values reported for AutoBVA are averages from 20 runs with very low standard deviation, extending or repeating its search while preserving previously found solutions is unlikely to significantly alter its overall diversity results.

When examining changes in quality and diversity across different SUTs, the extent of improvement varies across methods. \texttt{bytecount} achieves near-maximal quality early on for the majority of methods. Diversity is also relatively high even after 30 seconds, with uniform random and curiosity-based exploration being nearly ideal, while other methods continue to improve in RAC at 600 seconds. Both \texttt{bmi} and \texttt{circle} exhibit steady improvements in both quality and diversity as the time budget increases, showing that extended execution time benefits most methods in these SUTs. In contrast, the increased time budget for \texttt{date} primarily leads to a noticeable increase in diversity, while RPD remains relatively low.

A similar pattern can be observed for the Julia Base functions (Table~\ref{tab: juliabase_rac_rpd}), where AutoBVA maintains high RPD but consistently shows low RAC across all SUTs. In comparison, SETBVE configurations --- especially those combining the Sampler and Explorer --- achieve gains in diversity over time while maintaining reasonable quality. The addition of the Tracer component brings only minor changes to overall trends, indicating that the Explorer accounts for the majority of behavioral space exploration.

These variations indicate that while extended time budget generally enhances results, the degree of improvement depends on the characteristics of the SUT and the method used.

\begin{tcolorbox}[colback=blue!5, colframe=black!80, title=Key Findings (RQ1) - Quality and Diversity of Identified Boundaries]
\begin{itemize}
    \item SETBVE exploration methods improve diversity with extended execution, while the improvement in quality varies depending on the SUT and the framework configuration.
    \item Even the simplest SETBVE configuration (using only the CTS with bituniform sampling) often outperforms most tested methods in diversity while maintaining quality of boundary candidates.
    \item AutoBVA quickly achieves high quality but has low diversity and shows little improvement over time. 
    \item Random sampling is the weakest method, demonstrating minimal or no gains in either quality or diversity, even with a longer execution time.

\end{itemize}
\end{tcolorbox}

\subsection{RQ2 - Coverage of Behavioral Space}

In RQ2, we examine how different methods discover boundary candidates with distinct \textit{behaviors}, i.e., archive cells defined through behavioral descriptors. Therefore, we compare archive cells \textit{uniquely} found by specific methods across all their runs. Since differences among exploration strategies become clearer over longer runtimes, we analyze results from the 600-second runs. 

\begin{figure}
    \centering
    \includegraphics[width=\textwidth]{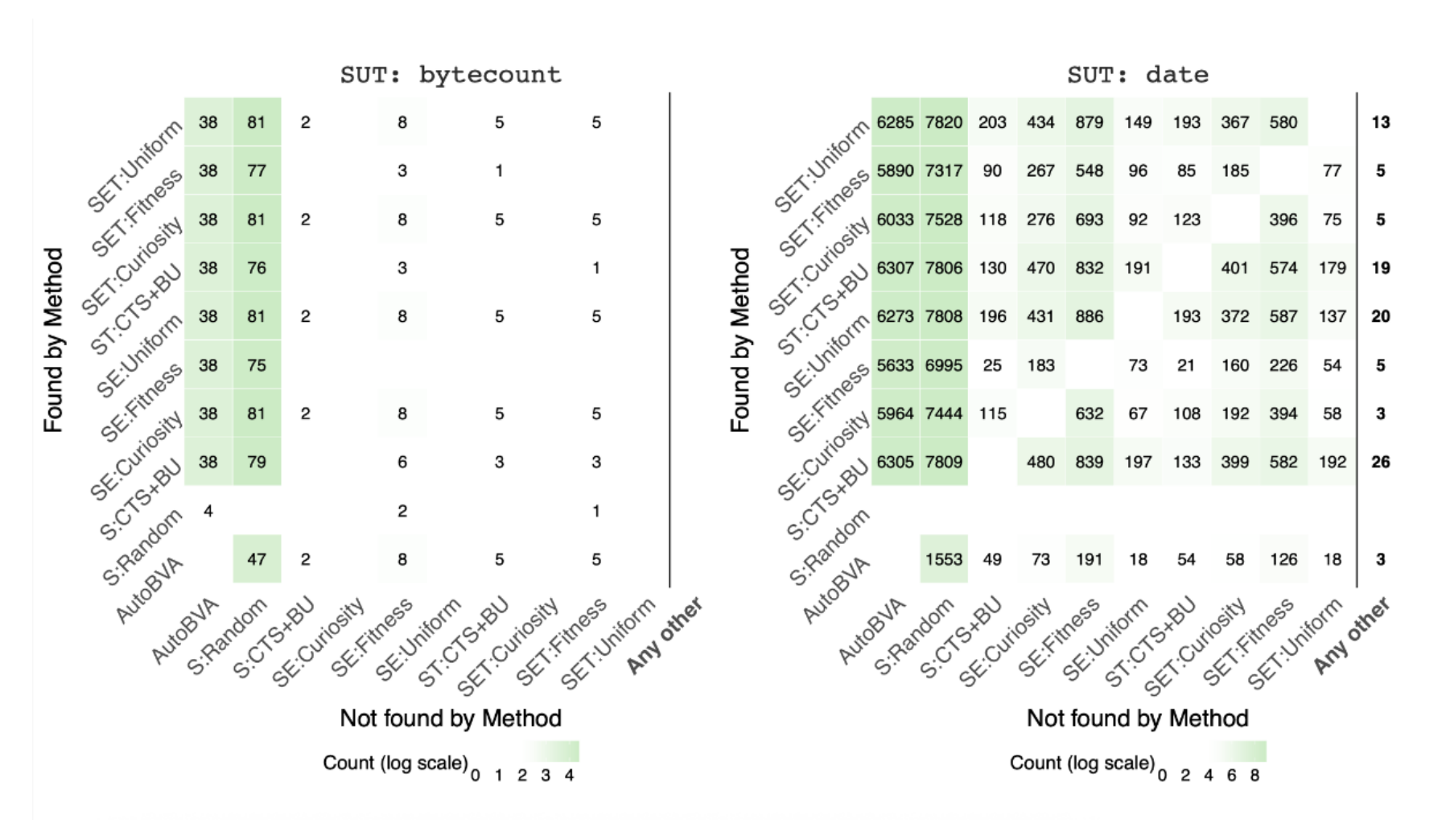}
    \caption{Pairwise comparison of uniquely found archive cells. The numbers represent unique archive cells discovered by the row method but missed by the column method. The color intensity corresponds to the logarithm of the count.}
    \label{fig:pairwiseheatmap} 
\end{figure}

Figure \ref{fig:pairwiseheatmap} provides a pairwise comparison of search strategies for the \texttt{bytecount} and \texttt{date} SUTs, chosen due to their contrasting complexities. First, we identify which search strategies (rows) have found more unique boundaries than \textit{any other} strategy (column), then we compare pairs of strategies. For instance, for \texttt{date}, SET-Uniform has found 13 boundaries that no other search found, but 149 boundaries that SE-Fitness could not find.

Our results show that the number of distinct archive cells identified by each method increases with SUT complexity. For \texttt{bytecount}, there is significant overlap between methods, while more complex SUTs such as \texttt{date} lead methods to discover mostly unique cells. The \texttt{bmi} and \texttt{circle} SUTs exhibit intermediate levels of uniqueness. The total count of unique cells ranges from 0–81 for \texttt{bytecount}, 0–620 for \texttt{circle}, 0–1326 for \texttt{bmi}, and up to 0–7820 for \texttt{date}.

For the \texttt{bytecount} SUT, the most unique behaviors are found in curiosity (SET-Curiosity, SE-Curiosity) and uniform random exploration (SET-Uniform, SE-Uniform). In contrast, fitness-based selection strategies find fewer unique cells among SETBVE configurations. Although AutoBVA identifies several unique cells, it overlooks many behaviors that other methods capture, as seen in the AutoBVA column in Figure~\ref{fig:pairwiseheatmap}.

For the \texttt{circle} and \texttt{bmi} SUTs, AutoBVA identifies several unique behaviors but misses many behaviors that SETBVE discovers, particularly those found by uniform random exploration and bituniform sampling. Within SETBVE, bituniform sampling alone performs effectively, but combining it with Explorer or Tracer reduces the number of unique archive cells found. Among exploration strategies, uniform random exploration generally identifies the most unique behaviors, followed closely by curiosity-based exploration, while fitness-based selection and random sampling are less effective. The addition of the Tracer component does not significantly enhance the number of unique behaviors and can even slightly reduce it for the \texttt{bmi} SUT.

Finally, for the \texttt{date} SUT, S-Bituniform (26), SE-Uniform (20), and ST-Bituniform configurations uncover the most distinct behaviors. AutoBVA finds a few unique behaviors (3) but fails to discover many cells found by other methods. Random sampling only identifies cells already discovered by other methods, offering no unique contributions. Nonetheless, comparing the pairwise portion of the heatmap and the unique behavior found (last column), we see that most behaviors are covered by multiple methods.

Table \ref{tab:example_bc} shows manually selected examples of \textit{archive cells}, and corresponding boundary candidates, to illustrate the type of boundary behavior that SETBVE discovered, but AutoBVA did not. We chose these examples based on their validity groups, RPD values, and the overall diversity of boundary candidates. The ``Cell'' column identifies the set of values from each behavioural descriptor. Inputs and their corresponding outputs are shown side by side.

\begin{table}[h]
    \centering
    \renewcommand{\arraystretch}{1.2}
    \caption{Examples of archive cells and their corresponding boundary candidates that were found by most SETBVE configurations but missed by AutoBVA, across the four SUTs: \textbf{bytecount}, \textbf{bmi}, \textbf{circle}, and \textbf{date}. Exception abbreviations: BErr (BoundsError), DErr (DomainError), AErr (ArgumentError), oor (out of range).}
    \label{tab:example_bc}
    \scriptsize 
    \arrayrulecolor{gray} 
    \begin{tabularx}{\textwidth}{lX X X X l}
        \toprule
        \textbf{Cell} & \textbf{Input 1} & \textbf{Input 2} & \textbf{Output 1} & \textbf{Output 2} & \textbf{Validity group} \\
        \midrule
        \multicolumn{6}{l}{\textbf{\texttt{bytecount}}} \\ 
        \noalign{\vskip 2.5pt}
        $[4, 0, 0, 0]$ & \texttt{-1} & \texttt{-2} & -1B & -2B & Valid - Valid \\
        $[22, 0, 0, 0]$ & \texttt{37949999999} & \texttt{37950000000} & 37.9 GB & 38.0 GB & Valid - Valid \\
        $[30, 0, 0, 0]$ & \texttt{-99999999999989} & \texttt{-99999999999990} & -99999999999989B & -99999999999990B & Valid - Valid \\

        \noalign{\vskip 3pt}
        \cdashline{1-6}[0.8pt/2pt]
        \noalign{\vskip 3pt}
        
        \multicolumn{6}{l}{\textbf{\texttt{circle}}} \\ 
        \noalign{\vskip 2.5pt}
        $[11, 0, 9, 0]$ & \texttt{(-79, -9)} & \texttt{(-80,-10)} & in & out & Valid - Valid \\
        $[6, 0, 5, 1]$ & \texttt{(-1, -1)} & \texttt{(0, 0)} & in & DErr(``Origin'') & Valid - Error \\
        $[5, 0, 6, 1] $ & \texttt{(1, 80)} & \texttt{(0, 0)} & out & DErr(``Origin'') & Valid - Error \\
        
        \noalign{\vskip 3pt}
        \cdashline{1-6}[0.8pt/2pt]
        \noalign{\vskip 3pt}
        
        \multicolumn{6}{l}{\textbf{\texttt{bmi}}} \\ 
        \noalign{\vskip 2.5pt}
        $[7, 0, 16, 0]$ & \texttt{(63, 9)} & \texttt{(63, 10)} & Normal & Obese & Valid - Valid \\
        $[6, 0, 12, 0]$ & \texttt{(36, 2)} & \texttt{(36, 3)} & Underweight & Overweight & Valid - Valid \\
        $[7, 0, 7, 1]$ & \texttt{(23, 1)} & \texttt{(23, -1)} & Normal & DErr(``Neg. input'') & Valid - Error \\

        \noalign{\vskip 3pt}
        \cdashline{1-6}[0.8pt/2pt]
        \noalign{\vskip 3pt}

        \multicolumn{6}{l}{\textbf{\texttt{date}}} \\ 
        \noalign{\vskip 2.5pt}
        $[13, 3, 0, 0]$ & \texttt{(3, 3, -1000)} & \texttt{(3, 3, -999)} & -1000-03-03 & -0999-03-03 & Valid - Valid \\
        $[15, 0, 33, 1]$ & \texttt{(31, 12, -100)} & \texttt{(32, 12, -99)} & -0100-12-31 & AErr(``Day: 32 oor'') & Valid - Error \\
        $[14, 1, 0, 2]$ & \texttt{(2246, 13, 0)} & \texttt{(2246, 12, 0)} & AErr(``Month:13 oor'') & AErr(``Day:2246 oor'') & Error - Error \\
        
        \bottomrule
    \end{tabularx}
    \begin{tablenotes}
        \footnotesize
        \item Cell coordinates: total input length, input length variance, output length difference (or abstraction number), number of exceptions.
    \end{tablenotes}
\end{table}

In the \texttt{bytecount} SUT, natural boundaries include transitions such as from kilobytes to megabytes within the VV group. As shown in Figure~\ref{fig:pairwiseheatmap}, this SUT exhibits the highest overlap between strategies in terms of discovered archive cells. Both AutoBVA and SETBVE successfully identified key transitions, including \texttt{kB} to \texttt{MB} (e.g., 999949 to 999950, yielding 999.9 kB to 1.0 MB), \texttt{MB} to \texttt{GB} (e.g., 999949999 to 999950000, or 999.9 MB to 1.0 GB), and others. Both methods also captured VE validity group boundary candidates, such as the transition from 1000.0 EB to a \texttt{BoundsError} (e.g., 999999999999994822656 to 999999999999994822657). SETBVE found some additional transitions, such as 54949999 to 54950000 (54.9 MB to 55.0 MB) and 37949999999 to 37950000000 (37.9 GB to 38.0 GB). Another example is the pair -99999999999989 to -99999999999990, which is still categorized as a boundary candidate under our current definition --- close input values leading to distinct program behaviors. Since we define program behavioral distinction based on the output distance between two outputs, even slightly different outputs like -99999999999989B and -99999999999990B qualify as a boundary. Overall, both methods discovered similar types of boundary candidates for this SUT.

For the \texttt{circle} SUT, both AutoBVA and SETBVE found adjacent boundary candidates such as DomainError (exception at the origin) to ``in'', and ``in'' to ``out'' (see Figure \ref{fig:trace_circle} for a visualization of the boundaries). However, SETBVE aims to maximize diversity across archive dimensions, including the output abstraction number used for the \texttt{circle} SUT. This objective allows SETBVE to also capture transitions between non-adjacent regions, provided these transitions represent distinct behavioral changes. Consequently, SETBVE discovered an additional boundary candidate transitioning directly from ``out'' to DomainError. Specifically, SETBVE identified inputs (1, 80)\footnote{The circle has a radius of 80 and is centered at the origin (0, 0).} and (0, 0) as the closest pair triggering this behavioral shift.

For the \texttt{bmi} SUT, we observe a similar pattern to the \texttt{circle} SUT in how AutoBVA prioritizes boundary candidates. Several non-adjacent categories (e.g.,DomainError to ``Overweight'', ``Underweight'' to Obese'') were not discovered by AutoBVA (see Figure~\ref{fig:trace_bmi} for a visualization of the boundaries). Although AutoBVA can find some non-adjacent categories, it typically detects only those where the input pairs lie very close together in the input space. Examples include DomainError to ``Severely Obese'' at (0, -1) and (0, 0), and ``Underweight'' to ``Severely obese'' at (9, 0) and (9, 1).

For the \texttt{date} SUT, the outputs are not categorical as in \texttt{circle} or \texttt{bmi}, but meaningful boundaries still exist based on calendar rules. Valid day values range from 1 to 31 depending on the month, and for February, the valid range also depends on whether the year is a leap year. Both AutoBVA and SETBVE successfully identified VE boundaries between day 0 and day 1 for most months. However, differences emerge when considering VE boundaries at the end of each month. Within the given time budgets, neither method found all possible end-of-month transitions, but SETBVE discovered more than AutoBVA. For example, SETBVE detected transitions from valid to invalid dates such as December 31 to December 32, and similar transitions in January, July, and April (30 to 31), which AutoBVA missed. In contrast, AutoBVA found end-of-month VE transitions that SETBVE did not, including those in March, June, and October. Both methods found several examples involving leap years (VE boundary candidates between February 28 and 29). The remaining end-of-month VE transitions --- in August, September, and November --- were identified by both methods. 

Overall, no single search strategy dominates in covering unique behaviours, and each approach has complementary strengths. As SUT complexity increases, the differences between methods become more pronounced, leading to larger gaps in coverage.

\begin{tcolorbox}[colback=blue!5, colframe=black!80, title=Key Findings (RQ2) - Coverage of Behavioral Space]
\begin{itemize}
    \item No single method consistently outperforms the others; among SETBVE configurations, curiosity-based and uniform random exploration typically discover more behavioral regions not found by other methods than fitness-based exploration.
    \item AutoBVA identifies some unique boundary behaviors but generally discovers fewer than SETBVE configurations, especially for complex SUTs and non-adjacent cases.

\end{itemize}
\end{tcolorbox}

\subsection{RQ3 - Tracing Identified Boundaries}

In RQ3, we investigate whether the Tracer can follow boundaries initially discovered by the Sampler and/or Explorer. The main objective of the Tracer is to enhance boundary refinement by exploring the vicinity of identified boundary candidates and locating additional ones. We evaluate the Tracer through visualizations of boundaries \footnote{Developing specific metrics to quantify tracing quality is left for future work.}.

We do not investigate tracing for the \texttt{bytecount} SUT because it has only a single input, resulting in a one-dimensional input space. In such a space, a boundary consists of isolated points rather than continuous regions, leaving no meaningful path to follow or trace.

\begin{figure}[H]
    \centering
    \includegraphics[width=0.8\textwidth]{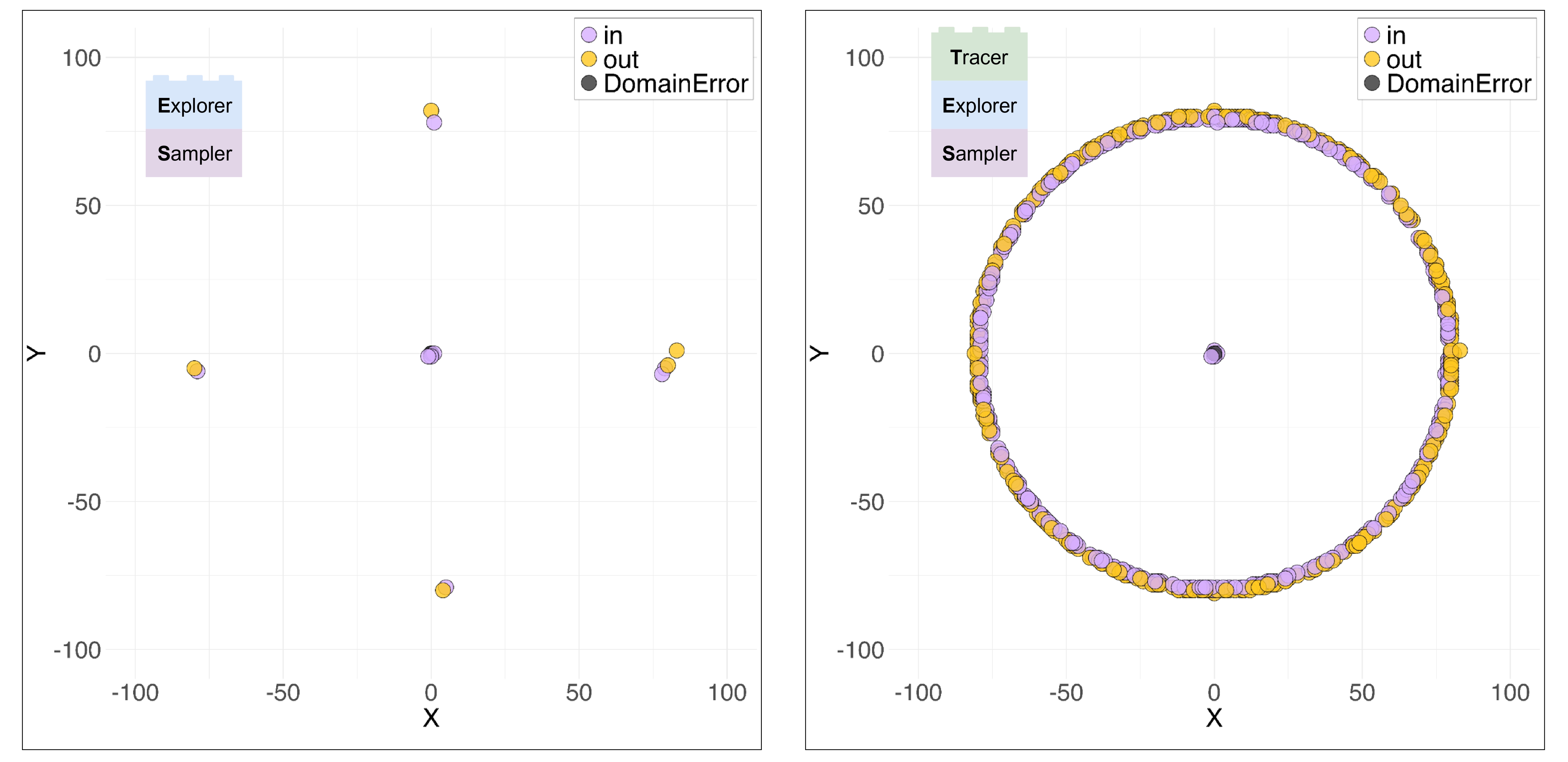}
    \caption{Example of boundary refinement for the \textbf{\texttt{circle}} SUT before and after applying the Tracer (600-second run with curiosity-based exploration).}
    \label{fig:trace_circle} 
\end{figure}

\begin{figure}[H]
    \centering
    \includegraphics[width=0.8\textwidth]{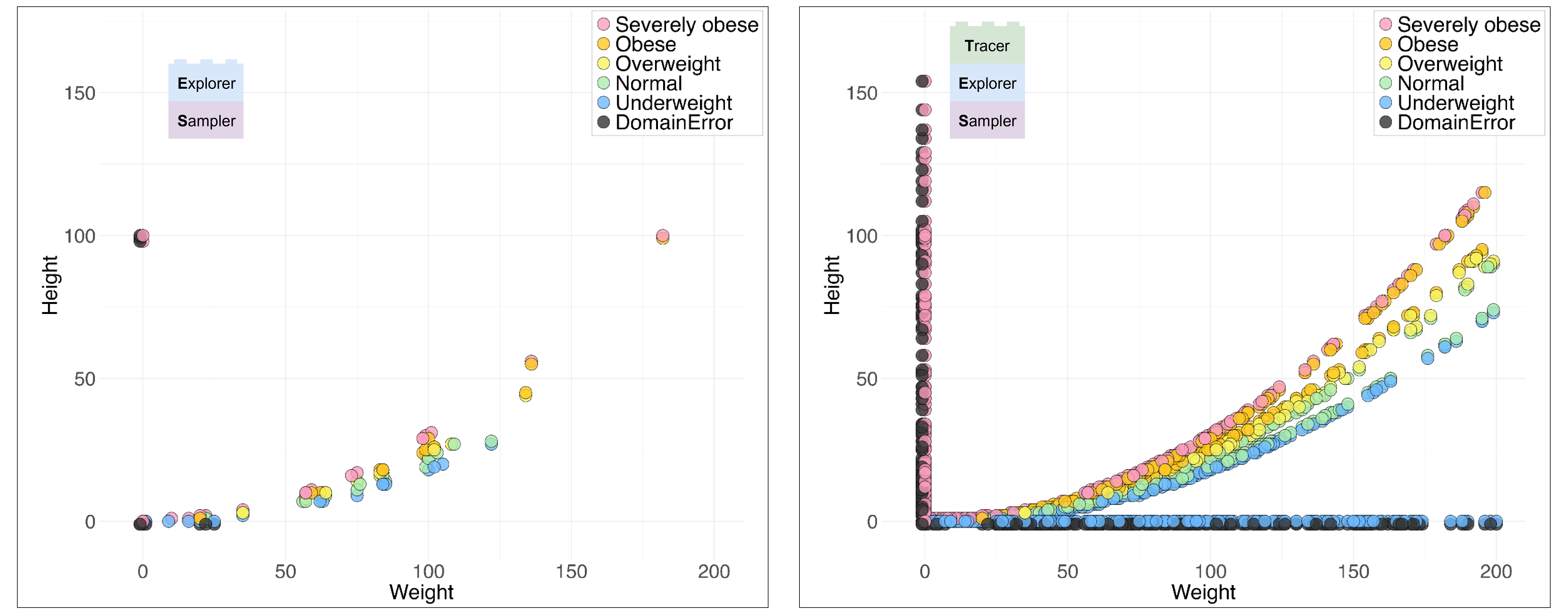}
    \caption{Example of boundary refinement for the \textbf{\texttt{bmi}} SUT before and after applying the Tracer (600-second run with fitness-based exploration).}
    \label{fig:trace_bmi} 
\end{figure}

In contrast, the \texttt{circle} and \texttt{bmi} SUTs serve as effective examples for boundary tracing due to their two-dimensional input spaces, where varying one or both input arguments naturally can lead to successful tracing and the discovery of additional boundary candidates. Figures \ref{fig:trace_circle} and \ref{fig:trace_bmi} illustrate this process. The left side of each figure shows boundary candidates initially discovered by the Sampler and Explorer, while the right side highlights the improved coverage achieved by the Tracer. This refinement better defines boundary shapes by revealing previously undetected parts of a boundary.

In the \texttt{bmi} SUT (Figure~\ref{fig:trace_bmi}), we observe that boundaries appear more concentrated at lower height and weight values, while transitions become more spread out as input values increase. Although this pattern may be specific to this SUT, it highlights how input characteristics can influence the distribution of behavioral transitions in the input space.

\begin{figure}[H]
    \centering
    \includegraphics[width=0.8\textwidth]{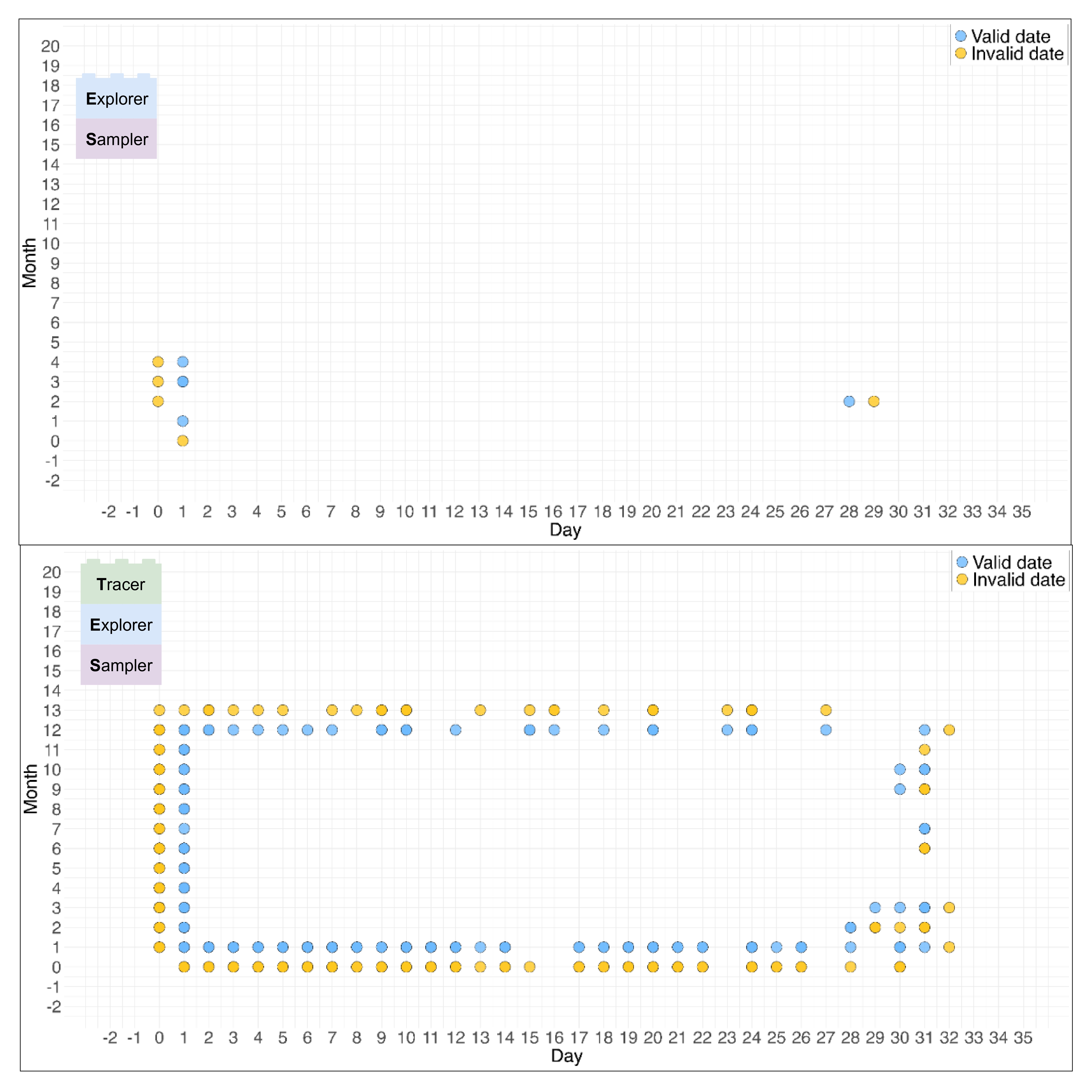}
    \caption{Example of boundary refinement for the \textbf{\texttt{date}} SUT before and after applying the Tracer (600-second run with uniform random-based exploration). The visualization omits the year input, meaning that the displayed day-month combinations belong to different years.}
    \label{fig:trace_date} 
\end{figure}

Boundary tracing for the \texttt{date} SUT is challenging because input arguments (day, month, and year) depend on each other. Ideally, tracing would involve varying one argument at a time—for example, adjusting only the year to test leap years or changing day and month to explore end-of-month transitions within a fixed year. However, the current Tracer implementation cannot fix specific inputs, limiting its precision. Despite this, the Tracer effectively finds additional input pairs close to previously identified boundaries, resulting in a denser mapping of behavioral transitions (see Figure~\ref{fig:trace_date}). In the figure, the left plot shows boundaries identified by the Explorer alone, while the right plot demonstrates improved completeness after applying the Tracer. The visualization excludes the year to highlight transitions between days and months.

For the \texttt{date} SUT, the Explorer alone captures a sparse and incomplete set of transitions, leaving many gaps in the mapping of behavioral changes. In contrast, the Tracer expands the set of identified transitions, revealing well-defined shifts --- particularly at the edges of valid date ranges. We observe clear transitions such as day 0 to day 1 and day 31 to day 32, along with a denser distribution of boundary candidates across multiple months. Some gaps remain, but the improvement towards completeness and clarity of these transitions highlights the potential of tracing, even in its initial form.

In summary, depending on SUT characteristics, visualizing boundary patterns helps to identify regions where program behavior changes more suddenly. Since it is often difficult to predict how boundaries are distributed in the input space, these visualizations provide valuable insights for deciding which input regions to prioritize during testing.

\begin{tcolorbox}[colback=blue!5, colframe=black!80, title=Key Findings (RQ3) - Tracing Identified Boundaries]
\begin{itemize}
    \item The Tracer component can extend the initial set of boundary candidates by exploring surrounding input and behavioral spaces, revealing additional boundary transitions and exposing previously undetected sections of the boundaries.

\end{itemize}
\end{tcolorbox}

%% file: files/discussion.tex
\section{Discussion}
\label{sec:discussion}

This study demonstrates that integrating Quality-Diversity optimization into Boundary Value Exploration yields a framework, SETBVE, that systematically uncovers a broader range of boundary behaviors than existing methods. In our experiments across ten SUTs, SETBVE maintained high-quality boundary candidates while significantly improving behavioral coverage compared to the baseline AutoBVA technique. For instance, with the \texttt{date} SUT, SETBVE configurations achieved approximately 52-80\% relative archive coverage compared to AutoBVA's 5.47\%, and similar patterns were observed across other tested systems. These results reveal that beyond maximizing a single boundary metric (program derivative), explicitly encouraging diversity across input\slash output descriptors enables the discovery of boundary regions that might otherwise remain undetected. This finding contributes to the field of software testing, where comprehensive boundary detection could potentially help identify failure points that might be missed by more focused testing approaches.

Our experiments reveal an inherent trade-off between the quality and diversity of boundary candidates. Maximizing behavioral diversity and identifying promising behavioral regions across different validity groups (VV, VE, EE) can offer a more comprehensive understanding of a system’s behavioral transitions. However, the SETBVE setup also identifies boundary candidates that span non-adjacent program domains by maximizing diversity across defined archive dimensions, challenging the conventional assumption that boundary candidates must originate from adjacent domains. We argue that this approach can be particularly valuable in high-dimensional input spaces where adjacency can become ambiguous or multi-dimensional. For instance, in a system under test that processes graphs to compute the number of strongly connected components, substantial changes to the graph structure may be required to alter the output. In such cases, an overemphasis on input adjacency — here, graph adjacency — could obscure significant behavioral transitions.

Our experiments reveal the quality-diversity trade-off in a more practical and tangible way. When a method like SETBVE emphasizes diversity, it inevitably diverts some computational effort from pure quality optimization. In practice, achieving the same quality as a focused approach like AutoBVA may require longer runtimes — or, somewhat unexpectedly, the impact on testing effectiveness may be minimal. We observed the latter: while SETBVE benefits from longer runtimes, it has to be considered that many boundary-pair candidates with the highest PD values are structurally similar, inflating quality without revealing genuinely novel behaviors. Meanwhile, SETBVE can uncover regions with locally high but globally lower PD values that a quality-focused method might overlook. Quantifying these effects with simple metrics remains challenging, suggesting that future work should involve real testers, ideally in interactive settings, to explore how best to balance diversity and quality for practical impact.

The quality impact of this trade-off is also complex to assess from a more subtle perspective, as there exists a theoretical limit to how many archive cells contains pairs that reach the highest PD levels. Our normalization of PD to RPD values makes the quality metrics more comparable across different regions of the behavioral space. Given our results, particularly a typical decrease in RPD for SETBVE for SUTs with more complex behavior (\texttt{date} and \texttt{bmi}), we expect there to be value in a future extension of SETBVE with within-cell optimization. Such an approach would not only identify more archive cells with diverse behavior but also find the locally optimal candidate within each cell. This within-cell refinement could be combined with the Tracer component, which already attempts to find boundary pairs in adjacent regions, to create a more comprehensive boundary exploration strategy that balances both diversity and quality across the behavioral space.

The Tracer component uncovers boundary candidates adjacent to those already identified. Although its impact on the metrics (RAC and RPD) is modest, visualizing it's effect reveals expanded regions of rapid change and subtle boundary patterns that our metrics doesn't capture. This discrepancy arises because the Tracer typically discovers additional high‐value boundary pairs that nonetheless map into existing archive cells, so RAC and RPD remain largely unchanged. In effect, the Tracer operates at a finer granularity than the archive, uncovering nuances that coarser cell-based metrics cannot detect. Future work should consider developing new metrics that capture these more fine-grained gains.

Given that both SUTs and testing contexts vary considerably, SETBVE's flexibility represents an important aspect of its design. The framework supports customization of key components in the QD approach, most notably the distance functions used to calculate quality (PD) and how behavioral diversity is defined through descriptors. The choice of behavioral descriptors directly shapes the search process and influences which boundary candidates are identified, but the behavioral descriptors can also be adjusted to meet domain-specific needs. Although this study primarily relied on generic input\slash output characteristics, the framework supports more tailored descriptors. Even with the simplest configuration, using only the Sampler component, the archive enables the generation of diverse candidates, as demonstrated by the results in RQ2. Future work should consider a wider set of choices for the distance and behaviora descriptor functions and how to exploit them for practical benefits.

\subsection{Limitations and Validity Threats}
While we have addressed several threats to validity, certain limitations warrant attention. Our measurement choices could affect the construct validity of our evaluation. Although Jaccard distance effectively captures string variations, other distance measures might yield different program derivative values, and thus could affect the quality diversity trade-off achieved.
To ensure that our choice of Jaccard distance did not artificially inflate SETBVE’s diversity advantage, we conducted a sensitivity analysis using the same distance function (string length) which was the default optimized by AutoBVA. Even under this alternative distance measure, SETBVE maintained significantly higher relative archive coverage (RAC) than AutoBVA. This confirms that the higher diversity of behaviors identified by SETBVE cannot be explained by using a more fine-grained (Jaccard) output distance function and thus mitigate this as an alternative explanation to our findings.

When it comes to the relative archive coverage (RAC) measure we used, it is based on empirical observations rather than establishing theoretical maximums on the number of possible archive cells. We try to mitigate this imperfection by aggregating results (and counting archive cells) over all runs, following accepted practices in related work. However, given that the input, output, and behavioral spaces studied are very large the runs might overlap less than can be expected, which might lessen the value of the normalization. Still, in practice, it will be very difficult to calculate, for a given SUT, how many cells can be covered making our normalization a reasonable and pragmatic choice.

Regarding internal validity, we controlled for biases in archive initialization by repeating each experiment 20 times and using sampling strategies proven effective in previous studies. However, the Tracer component's current implementation, which defines search regions based solely on input space rather than behavioral space, represents an area for refinement. For external validity, our focus on ten SUTs with integer inputs at the unit level limits generalizability to more complex software systems and input types. While our results suggest potential benefits for software testers, we have not empirically studied the impact on actual testing practices or outcomes, which limits our ability to make definitive claims about practical benefits.

\subsection{Future work}
Future research should build upon this study by addressing several interconnected areas. First, while SETBVE prioritizes diversity, further investigation into optimization techniques would enhance solution quality within regions where boundaries are already identified. The intra-cell optimization approach discussed earlier represents a promising direction. Second, extending SETBVE to accommodate different data types would overcome current limitations and improve applicability across diverse systems. Third, refining the tracing process to move beyond the basic approximations used in the current implementation would enable more precise descriptions of boundary patterns. Developing robust quantitative metrics for evaluating boundary tracing effectiveness would support more systematic assessment. Studies examining how testers interact with and benefit from the approach would provide valuable evidence for its practical utility. Investigating how SETBVE could be integrated into existing testing workflows would also enhance its applicability.


%% file: files/conclusion.tex
\section{Conclusion}
\label{sec:conclusion}

SETBVE introduces a novel approach to black-box Boundary Value Exploration (BVE) by integrating Quality-Diversity (QD) optimization to uncover a broader spectrum of behavioral transitions while maintaining reasonable quality. By combining the Sampler, Explorer, and Tracer components in a modular design, SETBVE adapts to different SUTs and testing objectives, enabling flexible exploration strategies. 
To evaluate the effectiveness of integrating QD into BVE, we proposed two novel metrics: Relative Program Derivative, to assess the relative boundariness (i.e., quality) of boundary candidates, and Relative Archive Coverage, to measure the diversity of solutions when using a grid-structured archive.

Our experiments demonstrate that SETBVE balances exploration and exploitation, revealing boundary behaviors that the baseline quality-focused method AutoBVA may overlook. The quantitative analysis across ten SUTs underscores SETBVE’s ability to discover diverse boundary candidates with only minor reductions in quantified quality. Qualitative analysis further illustrates how the Tracer component identifies additional behavioral transitions near existing boundaries, better highlighting regions of the input space with rapidly changing program behavior.

Notably, even the simplest SETBVE configuration --- employing only the Sampler --- performed comparably to more complex setups in both quality and diversity, suggesting that SETBVE’s core mechanisms, the grid-structured archive defined by general behavioral descriptors, effectively promote behavioral variety. However, optimizing solution quality within regions already populated with boundaries show room for improvement. Future work should refine the tracing process to provide more precise descriptions of boundary patterns and develop further metrics to assess the characteristics of discovered boundaries. Additionally, extending SETBVE to accommodate more complex data types will expand its applicability, while techniques to summarize and describe consecutive boundary pairs along boundaries could enhance the value and save time for testers.